\documentclass[12pt,draftclsnofoot,onecolumn]{IEEEtran}
\usepackage{graphicx,cite,epsfig,amssymb,amsmath,multirow,comment}
\usepackage{amsfonts}
\usepackage{mathrsfs}
\usepackage{indentfirst}
\usepackage{algorithm}
\usepackage{algorithmic}
\usepackage{setspace}
\usepackage{float}
\usepackage{bm}
\usepackage{ulem}
\usepackage{color}
\usepackage{amsmath}
\usepackage{amsthm}
\usepackage{stfloats}

\usepackage{fancyhdr}
\usepackage{url}
\usepackage{graphicx}
\usepackage{algorithm}
\usepackage{algorithmic}
\usepackage{amssymb}
\usepackage{array}
\usepackage{mathtools}
\usepackage{mathrsfs}
\usepackage{setspace}
\usepackage{multirow}
\usepackage{srcltx}
\usepackage{subfigure}
\usepackage{cite}
\usepackage{subeqnarray}
\usepackage{cases}
\usepackage{amssymb,amsmath,color,graphicx}
\usepackage{enumerate}
\allowdisplaybreaks

\theoremstyle{plain}

\theoremstyle{plain}
\newtheorem{prop}{\protect\propositionname}
\theoremstyle{plain}

\theoremstyle{plain}
\newtheorem{lem}{\protect\lemmaname}
\theoremstyle{remark}
\newtheorem{rem}[]{\protect\remarkname}

 \normalsize
\makeatother

\providecommand{\remarkname}{Remark}
\providecommand{\lemmaname}{Lemma}
\providecommand{\corollaryname}{Corollary}
\providecommand{\propositionname}{Proposition}
\providecommand{\theoremname}{Theorem}

\ifCLASSINFOpdf
\else
\fi

\allowdisplaybreaks

\begin{document}

\title{Cache-Aided Massive MIMO with Linear Precoding in Multi-cell Systems 
}
\author{Lin Xiang, Xiao Wei, Laura Cottatellucci, Robert Schober, and Tao Jiang \vspace{-3.2em}
\thanks{
This work has been presented in part at IEEE ICC, Shanghai, China, May 2019 \cite{Wei2019}.
}
}
\maketitle

\begin{abstract}
 \vspace{-0.3em}
In this paper, we propose a novel joint caching and massive multiple-input multiple-output (MIMO) transmission scheme, referred to as \emph{cache-aided massive MIMO}, for multi-cell downlink transmission to multiple cache-enabled receivers. With the proposed scheme, users who have cached (a portion of) the files that they request are offloaded and, hence, (partially) inactive during downlink transmission. The other users either benefit from the cache-enabled offloading for mitigating pilot contamination or exploit the cached but unrequested files to cancel interference during uplink channel estimation and downlink file reception. Moreover, by redesigning the transmit precoders based on the cache status of the users and channel state information, we gain additional degrees of freedom for massive MIMO transmission. For a given cache status, we analyze the equivalent content delivery rates (ECDRs), i.e., the average rates of delivering a requested file via both caching and massive MIMO transmission to the requesting user, for cache-aided massive MIMO employing re-designed maximum ratio transmission (MRT), zero-forcing (ZF) precoding, and regularized zero-forcing (RZF) precoding. Based on the derived results, the impact of (random) uncoded caching and coded caching on the performance of the re-designed precoding schemes is investigated. Simulation results validate our derivations and show that caching is beneficial for precoded downlink transmission as it enhances the transmit power allocation, mitigates intra- and inter-cell interference, and reduces the impairment caused by pilot contamination. Compared with conventional massive MIMO without caching and with cache-oblivious precoding, the proposed cache-aided massive MIMO scheme achieves a significantly higher ECDR even when the number of users approaches the number of transmit antennas. 
\end{abstract}


\vspace{-.8em}
\section{Introduction}
Massive multiple-input multiple-output (MIMO) is a key radio technology for fifth-generation (5G) wireless networks to improve spectral efficiency and cope with the explosive growth in cellular traffic and number of users \cite{Wong2017, N2013}. By employing a large number of antennas at the base station (BS), massive MIMO offers abundant spatial degrees of freedom and facilitates large multiplexing and diversity gains \cite{Lu2014, Marzetta2010}. The performance of massive MIMO with low-complexity linear precoding has been extensively studied in the literature. For single-cell massive MIMO systems, where the BS is equipped with $M$ antennas and communicates with $K$ single-antenna users, the performance has been shown to critically depend on the number of BS antennas per user, denoted by $\rho_0 \triangleq M/K$ \cite{Huh2011}. For example, when both $M$ and $K$ grow without bound while $\rho_0$ remains constant, the effective signal-to-interference-plus-noise ratio $\!$ (SINR) grows linearly with $\rho_0$ \cite{Yang2013}. Moreover, it was shown in \cite{Ngo2013,Raeesi2018} that linear precoding is asymptotically optimal, i.e., capacity-achieving, for massive MIMO systems with large $\rho_0$. However, impaired by pilot contamination and inter-cell interference, the performance of multi-cell massive MIMO may be limited even for large $M$ as long as $\rho_0$ is small \cite{Hoydis2013,Muller2014}. The authors of \cite{Hoydis2013} showed that $\rho_0$ has to be larger than $10$ to achieve $90\%$ of the optimal performance, obtained asymptotically for $\rho_0 \to \infty$. Besides, using subspace projection and power control in \cite{Muller2014}, the authors showed that the impact of pilot contamination decreases for large $\rho_0$.

However, due to emerging applications for smartphones and tablets, autonomous and automotive vehicles, and Internet-of-Things devices, the number of users in future wireless systems may grow significantly, even beyond the number of BS antennas \cite{Yu2017}, resulting in small $\rho_0$. In this case, conventional linear precoding based massive MIMO suffers from a significant performance loss. To improve the performance of massive MIMO systems with small $\rho_0$, researchers have been investigating advanced BS antenna systems such as extremely large aperture arrays \cite{Emil2019} and large intelligent reflecting surfaces \cite{Lu2014} for potential next-generation solutions. Considering that deploying a very large number of antennas may {{significantly increase hardware costs and processing complexity, this paper explores a user-side alternative for improving the performance of massive MIMO at low costs.}} In particular, we will show that wireless caching at the users can be exploited to enhance the capacity of massive MIMO in the small $\rho_0$ regime.

For many current and future applications, cache memory is expected to be available at the users' devices. By proactively pre-storing the most popular files in the users' caches during periods of low network traffic, fast access to these files is enabled without requiring over-the-air delivery and, at the same time, traffic congestion on backhaul links is ameliorated \cite{Semiari2018, Ng2018}. However, as the actual users' requests are not known during cache placement, the cached files may not be requested by the users later on, which imposes a fundamental limit on the performance gains of proactive user caching. To effectively utilize user caching, two approaches, which exploit advanced cache-enabled transmission to achieve additional performance gains, have been proposed in the literature \cite{Maddah-Ali2014, TWC22a, TWC22b, Xiang2018}. The first approach employs coded caching \cite{Maddah-Ali2014}. By carefully encoding the cached and delivered files, simultaneous multicast to multiple users is enabled such that each user can decode its requested file without suffering interference from other users \cite{Maddah-Ali2014}. However, for coded caching in wireless fading channels, the achievable rate of multicast is limited by the user with the worst channel conditions within the multicast group \cite{TWC22a}. Moreover, forming multicast groups based on the channel and cache status and the required decoding impose a large computational burden on the BS and users \cite{TWC22b}. An alternative approach, referred to as cache-aided non-orthogonal multiple access (NOMA) \cite{Xiang2018}, employs superposition coded broadcast transmission and exploits user's cached but unrequested files for canceling NOMA interference. By joint optimization of cache-enabled interference cancellation and successive interference cancellation, cache-aided NOMA can significantly improve the users' achievable rates, compared to conventional NOMA \cite{Xiang2018}. The idea of cache-aided NOMA has been recently combined with coded multicast transmission \cite{Dani2021} and extended to rate-splitting multiple access \cite{Jaafar2020}. However, when the number of users is large, the optimization of cache-aided NOMA becomes intractable.

In this paper, caching and massive MIMO are jointly designed to reap the benefits of both technologies while overcoming the high complexity of the techniques in \cite{Xiang2018}. Preliminary results for the proposed cache-aided massive MIMO scheme were reported in \cite{Wei2019}, where we assumed a single-cell system and perfect channel state information (CSI) at the BS. However, due to the impairment caused by noise, multiuser interference, and pilot contamination, practical CSI estimates are imperfect, which can severely limit the performance of linearly precoded massive MIMO transmission \cite{Marzetta2010,Hoydis2013,Muller2014}. This paper extends the design and analysis of the cache-aided massive MIMO scheme in \cite{Wei2019} to multi-cell systems with imperfect CSI knowledge at the BSs. We assume that each user is equipped with a cache memory and the BSs are equipped with a large number of antennas. To address the impact of imperfect CSI, we propose a novel cache-aided uplink channel estimation scheme and several cache-aided downlink precoding schemes for massive MIMO transmission. With the proposed schemes, the side information provided by cached files can be favorably exploited to enhance both uplink channel estimation and downlink data transmission, leading to significant performance gains for cache-aided massive MIMO in multi-cell systems. In particular, if a file cached at a certain user is later requested by the user itself, caching enables offloading of the transmission to this user. The  offloading does not only reduce the multiuser interference in both uplink channel estimation and downlink data transmission, but also increases the amount of transmit power allocable to those users that request uncached files. Additionally, if the files cached at a user are requested by other users, these files can still be exploited for interference cancellation at that user, avoiding the need for interference suppression via downlink precoding at the BSs. In this way, cache-aided massive MIMO frees up additional degrees of freedom for the transmission of the remaining files which further improves the performance. We note that synergies between caching and massive MIMO have also been exploited in \cite{Ngo2018,Papazafeiropoulos2018}. In \cite{Ngo2018}, several communication schemes combining coded caching with massive MIMO were proposed, which can enhance multicast transmission over fading channels and/or improve the spatial multiplexing capability of massive MIMO. Additionally, caching at the BSs and massive MIMO were jointly considered to enhance the performance of uplink communication in heterogeneous networks in \cite{Papazafeiropoulos2018}. However, these works did not exploit caching for channel estimation, i.e., reduction of pilot contamination, nor for interference mitigation.

The contributions of this paper can be summarized as follows:
\begin{itemize}
\item We propose a novel cache-aided massive MIMO scheme for improved channel estimation and data transmission in multi-cell systems with imperfect CSI. In addition to reaping the conventional advantages of caching and massive MIMO, the proposed scheme also increases the downlink achievable rate by exploiting caching for reducing pilot contamination and mitigating multiuser interference.

\item We redesign zero-forcing (ZF) and regularized zero-forcing (RZF) precoding at the BSs to jointly exploit the pilot decontamination and interference cancellation capabilities enabled by caching for enhanced data transmission. Different from conventional linear precoding techniques which require only CSI, the proposed linear precoders depend on both the CSI and the cache status. For arbitrary given cache status, we analyze the equivalent content delivery rates (ECDRs), i.e., the amount of requested content that can be delivered on average per unit time, for cache-aided massive MIMO when the BS employs maximum ratio transmission (MRT), ZF precoding, and RZF precoding.

\item We analyze the performance of the proposed precoding schemes for two widely adopted cache placement schemes, namely (random) uncoded caching and coded caching. Lower bounds on the ECDRs are derived in closed form. Our results reveal that the proposed precoding schemes can significantly improve the performance of both caching schemes irrespective of users' cache sizes.

\item Simulation results validate our analytical derivations and show that, compared with several baseline schemes, the proposed cache-aided massive MIMO scheme achieves a significantly higher ECDR, even when $\rho_0$ is small.
\end{itemize}

The remainder of this paper is organized as follows. In Section~II, the multi-cell system model and the proposed cache-aided massive MIMO scheme are introduced. We present the novel linear precoders and an analysis of their ECDRs in Section~III. In Section~IV, we investigate the impact of caching on the ECDR of the proposed scheme. In Section~V, the performance of the cache-aided massive MIMO scheme is evaluated by simulation. Finally, in Section~VI, we summarize our results and draw general conclusions. 

\emph{Notations}: In this paper, we use boldface capital and lower case letters to denote matrices and vectors, respectively. ${{\mathbb C}^{N_r \times N_t}}$ is the set of complex-valued $N_r \times N_t$ matrices.
${\bf{A}}^{\rm H}$, ${\bf{A}}^{\rm T}$, and $\mathrm{tr} ({\bf{A}})$ represent the complex conjugate transpose, 
transpose, and 
trace of matrix $\bf{A}$, respectively;
${\bf{A}}^{-1}$ is the inverse of a square matrix $\bf{A}$;
$[{\bf{A}}]_{k,m}$ is the element in the $k$th row and the $m$th column of matrix ${\bf{A}}$.
$\mathcal{CN} \left(0,\sigma^2\right)$ denotes the complex Gaussian distribution with zero mean and variance $\sigma^2$.
${\bf{I}}_N$ is the $N \times N$ identity matrix.
$\Pr(\cdot)$ and $\mathcal{E}\{\cdot\}$ are the probability and the expectation operators, respectively;
{{$\mathrm{diag} (\mathbf{x})$ is a diagonal matrix whose main diagonal elements are given by vector ${\bf{x}}$;}}
$\left\| {\bf{x}} \right\|$ and $\left|x\right|$ are the Euclidean norm of vector ${\bf{x}}$ and the absolute value of scalar $x$, respectively; $\left|\mathbb{X}\right|$ is the cardinality of set $\mathbb{X}$. $\mathbb{X}\times\mathbb{Y}$ and $\mathbb{X} \setminus \mathbb{Y}$ are the Cartesian product and the difference of sets $\mathbb{X}$ and $\mathbb{Y}$, respectively.
$A \rightarrow B$ indicates that $A$ converges to $B$ in the limit, and finally, $C_m^n$ is the binomial coefficient.
\vspace{-1em}

\section{System Model} \label{Cache-Aided Multi-cell Massive MIMO}
\subsection{Cache-Aided Multi-cell Massive MIMO System} \label{System Model}
We consider a massive MIMO system comprising $B$ cells. 
The BS in each cell is equipped with $M$ antennas and serves $K$ single-antenna users. Define sets $\mathbb{B} \triangleq \{1,\ldots,B\}$ and $\mathbb{K} \triangleq \{1,\ldots,K\}$. We then use tuple $(j,l)$ to denote user $l \in \mathbb{K}$ in cell $j \in \mathbb{B}$ and the user of interest is denoted by tuple $(b,k)$. We assume that a library of $L_s$ files with index set $\mathbb{L}_s \triangleq \{1,\ldots,L_s\}$ is available at each BS. Each file has a size of $F$ MBytes. To reduce the file downloading delay, each user is equipped with a cache memory of size $L_uF$ MBytes. We assume $L_u < L_s$, i.e., each user can only cache a portion of the library files, and a given file may be partially or fully cached. 

The system operates in two phases: a placement phase and a delivery phase. During the placement phase, each user receives $L_u$ files from the library and stores them into its own cache. For convenience, for the moment we assume that each of the $L_u$ files is entirely cached. The extension to the case of partial caching will be discussed in Section~\ref{Cache Placement Scheme}. The file placement is completed prior to the time of request, e.g., in the early mornings when cellular traffic is low. In the delivery phase, each user may request one of the library files. Let $c_{j,l,j',l'}=0$ if the file requested by user $(j,l)$ has been cached at user $(j',l')$ and $c_{j,l,j',l'}=1$ otherwise. In general, $c_{j,l,j',l'}$ and $c_{j',l',j,l}$ can have distinct values unless $(j,l)=(j',l')$. Using this notation, if a requested file is cached by the requesting user itself, i.e., if $c_{j,l,j,l}=0$, the file is fetched from the user's cache instantly. In this case, user $(j,l)$ is considered {\emph{inactive}} as it requires no cellular transmission. Otherwise, if $c_{j,l,j,l}=1$, the requested file has to be transmitted by BS $j$, and user $(j,l)$ is considered to be {\emph{active}}. Thus, the number of active users in cell $j$ is given by $\overline{K}_j=\sum\nolimits_{l \in \mathbb{K}} c_{j,l,j,l}$.

For file delivery, we assume that a time division duplex (TDD) protocol is employed and that the uplink and downlink channels are reciprocal. The channel coefficient between user $(j,l)$ and the $m$th antenna of BS $j'$, denoted by $h_{j,l,j',m}$, is modeled as
\begin{equation}\label{eq2.1}
h_{j,l,j',m} = g_{j,l,j',m}\sqrt{\beta_{{j,l,j'}}},
\end{equation}
where $g_{j,l,j',m}$ is the small-scale fading coefficient from the $m$th antenna of BS $j'$ to user $(j,l)$ and follows a complex Gaussian distribution $\mathcal{CN}(0,1)$. $\beta_{j,l,j'}$ models the pathloss and shadowing effects between user $(j,l)$ and BS $j'$. We assume that $\beta_{j,l,j'}$ remains constant over a large number of coherence time intervals such that its value can be accurately estimated and is known at BS $j'$ \cite{Zhu2016, Kay1993}. We define the fading vector from BS $j'$ to user $(j,l)$ as ${\bf{g}}_{j,l,j'} \triangleq \left[g_{j,l,j',1},\ldots,g_{j,l,j',M}\right]^{\mathrm T}$. Then, the channel matrix between  users in cell $j$ and BS $j'$, denoted as ${\bf{H}}_{j,j'}\in\!{{\mathbb C}^{{K} \times {M}}}$ with $[{\bf{H}}_{j,j'}]_{l,m}=h_{j,l,j',m}$, is given as
\begin{equation}\label{eq2.2}
{\bf{H}}_{j,j'} ={\bf{D}}_{j,j'}^{\frac{1}{2}} {\bf{G}}_{j,j'},
\end{equation}
where ${\bf{D}}_{j,j'}\triangleq\mathrm{diag}([\beta_{j,1,j'},\ldots,\beta_{j,K,j'}])\in\!{{\mathbb C}^{{K} \times {K}}}$ and ${\bf{G}}_{j,j'}=[{\bf{g}}_{j,1,j'},\ldots,{\bf{g}}_{j,K,j'}]^{\rm T} \!\in\!{{\mathbb C}^{{K} \!\times\! {M}}}$ is the matrix of fading coefficients between 
users in cell $j$ and BS $j'$ with $[{\bf{G}}_{j,j'}]_{l,m}=g_{j,l,j',m}$.
\vspace{-1em}

\subsection{Uplink Channel Estimation and Pilot Decontamination via Caching} \label{Uplink Channel Estimation}
Exploiting reciprocity, the downlink channels are estimated at the BSs based on pilot transmission in the uplink. Thereby, at the beginning of each coherence time interval, all active users simultaneously transmit mutually orthogonal pilot sequences comprising $\tau$ symbol intervals. Let $\sqrt{\tau}{\bf{x}}_l\in {{\mathbb C}^{\tau \times 1}}$ be the pilot sequence transmitted by user $(j,l)$ in cell $j$, where
\begin{equation}
{\bf{x}}_l^{\mathrm{H}}{\bf{x}}_{l'} = \begin{cases}
0, & \textrm{if } l\neq l', \\
1, & \textrm{otherwise}.
\end{cases}
\end{equation}
Inactive users neither require downlink date transmission nor emit pilots for channel estimation.
Taking into account the cache status of the users, the signal received at BS $j$, denoted by ${\bf{Y}}_{j}^{\mathrm {u}}\in\!{{\mathbb C}^{{\tau} \times {M}}}$, is given as follows
\begin{equation}\label{eq3.1}
{\bf{Y}}_{j}^{\mathrm {u}} =\sqrt{p\tau} \sum\nolimits_{(j',l) \in \mathbb{B} \times \mathbb{K}} c_{j',l,j',l}{\bf{x}}_l {\bf{h}}_{j',l,j}^{\mathrm{H}} + {\bf{N}}_{j}^{\mathrm {u}},\;\;\;j\in \mathbb{B},
\end{equation}
where $p$ is the power of a pilot symbol, ${\bf{h}}_{j',l,j}=[h_{j',l,j,1},\ldots,h_{j',l,j,M}]^{\mathrm{T}}\in {{\mathbb C}^{M \times 1}}$ is the unknown channel vector to be estimated, and ${\bf{N}}_{j}^{\mathrm {u}}\in\!{{\mathbb C}^{{\tau} \times {M}}}$ is an additive white Gaussian noise (AWGN) matrix whose mutually independent elements $[{\bf{N}}_{j}^{\mathrm {u}}]_{l,m}$ follow distribution $\mathcal{CN} \left(0,1\right)$.

Without loss of generality, we assume that the user of interest has not cached the file that it requests, i.e., $c_{b,k,b,k}=1$, and the file has to be transmitted by BS $b$. In Section~\ref{Cache Placement Scheme}, when partial caching is considered, we assume $c_{b,k,b,k}>0$ for the user of interest. A comprehensive study for all possible cache status is postponed to Section IV. Then, the minimum mean square error (MMSE) estimate of channel ${\bf{h}}_{b,k,j}$, denoted by ${\hat{\bf{h}}}_{b,k,j}$, is obtained at BS $j$ as follows
\begin{align}\label{eq3.2}
{\hat{\bf{h}}}_{b,k,j}^{\mathrm{H}} &=\sqrt{p\tau}\beta_{b,k,j}{\bf{x}}_{k}^{\mathrm{H}}\left( {\bf{I}}_{\tau} + {\bf{x}}_{k}\left(p\tau\sum\nolimits_{j'\in \mathbb{B}} c_{j',k,j',k}\beta_{j',k,j}\right){\bf{x}}_{k}^{\mathrm{H}}\right)^{-1}{\bf{Y}}_{j}^{\mathrm {u}},\;\;\;j\in \mathbb{B}.
\end{align}
Let ${\tilde{\bf{h}}}_{b,k,j}={\bf{h}}_{b,k,j}-{\hat{\bf{h}}}_{b,k,j}$ be the channel estimation error vector.
Owing to the MMSE estimation, ${\hat{\bf{h}}}_{b,k,j}$ and ${\tilde{\bf{h}}}_{b,k,j}$ are independent random vectors distributed as ${\hat{\bf{h}}}_{b,k,j} \sim \mathcal{CN} \left(0,\hat{\beta}_{b,k,j}{\bf{I}}_{M}\right)$ and ${\tilde{\bf{h}}}_{b,k,j} \sim \mathcal{CN} \left(0,\tilde{\beta}_{b,k,j}{\bf{I}}_{M}\right)$\cite{Zhu2016, Kay1993}, respectively, where
\begin{subequations}\label{eq5.0}
\begin{align}
\label{eq3.5}\hat{\beta}_{b,k,j}=\frac{p\tau\beta_{b,k,j}^2}{1+p\tau\left(\beta_{b,k,j}+\sum\nolimits_{j'\neq b} c_{j',k,j',k}\beta_{j',k,j}\right)}, \\
\label{eq3.6}\tilde{\beta}_{b,k,j}=\frac{(1+p\tau\sum\nolimits_{j' \neq b} c_{j',k,j',k}\beta_{j',k,j})\beta_{b,k,j}}{1+p\tau\left(\beta_{b,k,j}+\sum\nolimits_{j'\neq b} c_{j',k,j',k}\beta_{j',k,j}\right)}.
\end{align}
\end{subequations}

In \eqref{eq3.2}, pilot contamination arises when pilot sequence ${\bf{x}}_k$ is also reused in other cells, i.e., if $c_{j',k,j',k}=1$ for cell $j' \in \mathbb{B} \setminus \{b\}$, which degrades the accuracy of the channel estimation. Due to pilot contamination, the MMSE channel estimates of active users $(b,k)$ and $(j',k)$ for $j'\neq b$, are collinear, i.e., 
${\hat{\bf{h}}}_{j',k,j} = \frac{\beta_{j',k,j}}{\beta_{b,k,j}} {\hat{\bf{h}}}_{b,k,j}$.
Additionally, if user $(j',k)$ is inactive, i.e., $c_{j',k,j',k}=0$, it neither transmits pilot symbols nor causes interference to user $(b,k)$ during 
channel estimation. Hence, in \eqref{eq3.2}, cache-enabled offloading can be exploited to mitigate pilot contamination and improve the quality of 
channel estimates.
\vspace{-1em}

\subsection{Downlink Data Reception and Cache-Enabled Interference Cancellation} \label{Downlink Date Transmission}
Data transmission is initiated at the BSs upon the completion of channel estimation. The signal received by user $(b,k)$ and denoted by $y_{b,k}^{\mathrm{d}}$ is given as follows
\begin{equation}
\!y_{b,k}^{\mathrm {d}}\!=\! \sum\nolimits_{(j,l) \in \mathbb{B} \times \mathbb{K}} {\bf{h}}_{b,k,j}^{\mathrm H}{\bf{w}}_{j,l}s_{j,l}+ n_{b,k}^{\mathrm {d}}, \label{eq4.1}
\end{equation}
where $s_{j,l}$ is the transmit data symbol intended for user $(j,l)$ with $\mathcal{E} \{ \left| s_{j,l} \right| ^2 \}=E_{j,l}$, ${\bf{w}}_{j,l}\!\in\!{{\mathbb C}^{{M} \times {1}}}$ is the precoding vector for user $(j,l)$, and $n_{b,k}^{\mathrm {d}}$ is the AWGN at user $(b,k)$ following distribution $\mathcal{CN} \left(0,1\right)$.
We assume that the total transmit power of each BS is $E_0$. To satisfy the total transmit power constraint in cell $j$, we require $\sum\nolimits_{l \in \mathbb{K}} \!c_{j,l,j,l}E_{j,l} \! = \! E_0$ and $\mathcal{E} \{ \left\| {\bf{w}}_{j,l} \right\| ^2 \}=1$. 

For further analysis of the downlink data communication, we decompose the received signal in \eqref{eq4.1} as follows
\begin{equation}
\!y_{b,k}^{\mathrm {d}}\!=\!{\underbrace{{\hat{\bf{h}}}_{b,k,b}^{\mathrm H}{\bf{w}}_{b,k}s_{b,k}}_{\textrm{desired signal}}} \!+\! {\underbrace{\sum\limits_{l \neq k} \! c_{b,l,b,l}{\hat{\bf{h}}}_{b,k,b}^{\mathrm H}{\bf{w}}_{b,l}s_{b,l}}_{\textrm{intra-cell  interference}}}
\!+\!{\underbrace{\! \sum\limits_{(j,l) \in (\mathbb{B} \setminus \{b\}) \times \mathbb{K}} \!\!\!\!\!\!\!\!\!\!\!\! c_{j,l,j,l}{\hat{\bf{h}}}_{b,k,j}^{\mathrm H}{\bf{w}}_{j,l}s_{j,l}}_{\textrm{inter-cell interference}}} \!+\!\!\! {\underbrace{\sum\limits_{(j,l) \in \mathbb{B} \times \mathbb{K}}\!\!\!\!\!\! c_{j,l,j,l} {\tilde{\bf{h}}}_{b,k,j}^{\mathrm H}{\bf{w}}_{j,l}s_{j,l}}_{\textrm{interference due to CSI estimation error}}} \!\!\!\!+\! n_{b,k}^{\mathrm {d}}.\!\! \label{eq4.1.2}
\end{equation}
Herein, we have followed the approach in \cite{TM2016} to account for the impairment caused by pilot contamination. Moreover, as \cite{TM2016}, we do not use channel estimation in the downlink but leverage the channel hardening at the users. In \eqref{eq4.1.2}, the desired signal is impaired not only by the intra- and inter-cell interference caused by simultaneous transmissions to multiple users but also by the CSI estimation errors caused by pilot contamination.
However, if $c_{j,l,j,l}=0$, user $(j,l) \neq (b,k)$ is \emph{offloaded} for data transmission and hence, causes no interference. Additionally, if $c_{b,l,b,l}=1$ and $c_{b,l,b,k}=0$, i.e., the file requested by active user $(b,l)$ is cached at user $(b,k)$, this cached file can be exploited for \emph{intra-cell interference cancellation} \cite{Xiang2018} to improve downlink data reception. In particular, by re-encoding this cached file and subtracting the corresponding signal from $y_{b,k}^{\mathrm {d}}$, the intra-cell interference caused by user $(b,l)$ can be removed at user $(b,k)$. The idea of interference cancellation can also be extended to inter-cell interference. Thereby, if $c_{j,l,j,l}=1$ and $c_{j,l,b,k}=0$ with $j\neq b$, the inter-cell interference from user $(j,l)$ can be removed at user $(b,k)$ when the same coding and modulation schemes are adopted in all cells or the employed coding and modulation schemes are known at all users.

Hence, the interference signals that can be canceled at user $(b,k)$ by exploiting the cached files are given by $s_{j,l}$, $(j,l) \in \mathbb{I}_{b,k}$, where $\mathbb{I}_{b,k} = \{ (j,l)\in \mathbb{B}\times \mathbb{K} \mid c_{j,l,j,l}=1, c_{j,l,b,k}=0,(j,l)\neq (b,k) \}$. Thereby, each re-encoded signal $s_{j,l}$, $(j,l) \in \mathbb{I}_{b,k}$ is scaled by ${\bf{h}}_{b,k,j}^{\rm H}{\bf{w}}_{j,l}$ before being subtracted from the received signal in \eqref{eq4.1.2}. This cancellation requires knowledge of ${\bf{h}}_{b,k,j}^{\rm H}{\bf{w}}_{j,l}$, which can be estimated locally at user $(b,k)$ by treating the cached file as training sequence \cite[Sec. III-D]{Muller2014}, while no knowledge about the requests nor the cache status of the other users is needed. Since ${\bf{h}}_{b,k,j}^{\rm H}{\bf{w}}_{j,l}$ is a scalar value, the estimation error is negligible. Additionally, if $s_{j,l} = s_{j',l'}$ holds for $(j,l)\neq (j',l')$, i.e., when the interfering users $(j,l)$ and $(j',l')$ have requested the same file and this file is transmitted using the same modulation and coding scheme, the interference signals $s_{j,l}$ and $s_{j',l'}$ will be canceled together based on the estimate of the effective channel ${\bf{h}}_{b,k,j}^{\rm H}{\bf{w}}_{j,l} + {\bf{h}}_{b,k,j'}^{\rm H}{\bf{w}}_{j',l'}$.

We assume that the considered multi-cell system is synchronized and enables perfect interference cancellation. Consequently, by exploiting caching at user $(b,k)$, the only active users that cause interference to user $(b,k)$ are the ones in set $\mathbb{U}_{b,k} \triangleq \{ (j,l)\in \mathbb{B}\times \mathbb{K} \mid c_{j,l,j,l} = 1, c_{j,l,b,k} =1,(j,l)\neq (b,k)\}$. Define $\mathbb{V}_{b,k}=\mathbb{U}_{b,k}\cup \{(b,k)\}$, whereby $\mathbb{I}_{b,k} \cup \mathbb{V}_{b,k}$ gives the set of active users in the considered system.
Furthermore, we define $\mathbb{U}_{b,k,j}$ as the set of users in cell $j$ that cause interference to user $(b,k)$ such that $\mathbb{U}_{b,k}=\cup_{j \in \mathbb{B}}\mathbb{U}_{b,k,j}$. Then, $\mathbb{U}_{b,k,b}$ contains the set of interfering users in cell $b$, which is the set of active users other than user $(b,k)$. The residual received signal of user $(b,k)$ after interference cancellation, denoted by $y_{b,k}^{\mathrm{d,IC}}$, is given as follows
\begin{align}\label{eq4.2}
y_{b,k}^{\mathrm{d,IC}}\!=\!{\hat{\bf{h}}}_{b,k,b}^{\mathrm H}{\bf{w}}_{b,k}s_{b,k} \!+\!\!\sum\nolimits_{(j,l) \in \mathbb{U}_{b,k}} \!\!{\hat{\bf{h}}}_{b,k,j}^{\mathrm H}{\bf{w}}_{j,l}s_{j,l} \!+\!\!\sum\nolimits_{(j,l) \in \mathbb{V}_{b,k}}\!\! {\tilde{\bf{h}}}_{b,k,j}^{\mathrm H}{\bf{w}}_{j,l}s_{j,l}\! +\! n_{b,k}^{\mathrm {d}}.
\end{align}

\vspace{-1em}
\subsection{Equivalent Content Delivery Rate} \label{Equivalent Content Delivery Rate}
For evaluating the performance of user $(b,k)$, we adopt the ECDR defined as follows\footnote{When $L_d=0$ or $c_{b,k,b,k}=0$, the requested file can be fetched from the cache instantly, leading to $\mathcal{R}_{b,k} \rightarrow \infty$.} \cite{Ngo2018}
\begin{align}\label{eq5.0}
\mathcal{R}_{b,k}=\tfrac{F}{L_{d}}R_{b,k}.
\end{align}
Herein, $R_{b,k}$ is the ergodic achievable rate of user $(b,k)$ and $L_d$ is the length of the file to be delivered. We have $L_d=F$ for user $(b,k)$ if it has not cached the file that it requests. However, if a user has cached only a portion of the file that it requests, as in Section IV-B, we have $L_d < F$.
Observe that $L_d/R_{b,k}$ represents the time required to complete the delivery of the file requested by user $(b,k)$. Therefore, $\mathcal{R}_{b,k}$ characterizes the aggregate rate of content delivery via joint caching and massive MIMO transmission. Assume that the file delivery for each active user spans a large number of coherence time intervals. Then, for the proposed cache-aided massive MIMO, the ergodic achievable rate is given by
\begin{align}\label{eq5.2}
R_{b,k}=\mathcal{E}\left\{\log_2\left(1+\gamma_{b,k}\right)\right\},
\end{align}
where $\gamma_{b,k}$ is the (effective) SINR of user $(b,k)$ given by \cite{TM2016}
\begin{align}\label{eq5.1}
\gamma_{b,k}=\frac{P_{b,k}^{\mathrm{s}}} {\sum\nolimits_{(j,l) \in \mathbb{U}_{b,k}\!}\!P_{b,k,j,l}^{\mathrm{i}} \!+\!  \sum\nolimits_{(j,l) \in \mathbb{V}_{b,k}\!}\!P_{b,k,j,l}^{\mathrm{e}} \!+\!1}.
\end{align}
Here, $P_{b,k}^{\mathrm{s}}=E_{b,k}| {\hat{\bf{h}}}_{b,k,b}^{\mathrm H}{\bf{w}}_{b,k} |^2$, $P_{b,k,j,l}^{\mathrm{i}}=E_{j,l}|  {\hat{\bf{h}}}_{b,k,j}^{\mathrm H}{\bf{w}}_{j,l} |^2$, and $P_{b,k,j,l}^{\mathrm{e}}=E_{j,l}| {\tilde{\bf{h}}}_{b,k,j}^{\mathrm H}{\bf{w}}_{j,l} |^2$
are the power 
of the desired signal, the residual intra- and inter-cell interference signals, and the interference caused by the CSI estimation error, respectively.
Since $f(x)=\log_2(1+\frac{1}{x})$ is a convex function, by exploiting Jensen's inequality, a lower bound on the ECDR 
$\mathcal{R}_{b,k}$ is obtained as \cite{TM2016}
\begin{align}\label{eq5.3}
\mathcal{R}_{b,k}\geq \tilde{\mathcal{R}}_{b,k} \stackrel{\Delta}{=}\tfrac{F}{L_{d}}\log_2\left(1+\left(\mathcal{E}\left\{\gamma_{b,k}^{-1}\right\}\right)^{-1}\right).
\end{align}
In the following sections, this lower bound will be used to analyze the performance of cache-aided massive MIMO for different precoding and caching strategies.

\vspace{-0.5em}
\section{Cache-Aided Linear Precoder Design and Performance Analysis} \label{Cache-Aided Linear Precoder Design}
During data transmission, the benefits introduced by cache-enabled pilot decontamination and interference cancellation as discussed in Section~\ref{Cache-Aided Multi-cell Massive MIMO} can be further exploited for precoder design at the BSs, which is considered in this section. Assume that the cache status of the users is given and known at their associated BSs. 
We propose novel precoder designs for cache-aided massive MIMO that can intelligently adapt themselves according to both the estimated CSI and the users' cache status, rather than ignoring either of the available information. Throughout this work, we consider linear MRT, ZF, and RZF precoders, which are preferred in practical massive MIMO implementations for their low computational complexity \cite{Access21}. We suitably modify these precoding schemes to leverage the caching gains.

\vspace{-1em}
\subsection{Maximum Ratio Transmission} \label{Maximum-Ratio Transmission}
MRT precoding at the BS ensures that the signals transmitted over different BS antennas add up constructively at the intended user. Hence, if perfect CSI is available, MRT precoding maximizes the received signal power and thus, the signal-to-noise ratio (SNR). However, when $\rho_0$ is small, MRT suffers from severe multiuser interference, especially in the high SNR regime. Moreover, for the considered multi-cell massive MIMO system, the performance of MRT precoding further deteriorates due to imperfect CSI at the BSs. However, by exploiting user caching for interference cancellation and offloading, the performance of MRT can be improved in a straightforward manner. In particular, the MRT precoding vector of user $(b,k)$ is given by \cite{Yang2013,Chien2016}
\begin{equation}\label{eq6.1}
{\bf{w}}_{b,k}^{\mathrm{MRT}}=\sqrt{\lambda_{b,k}^{\mathrm{MRT}}} {\hat{\bf{h}}}_{b,k,b},
\end{equation}
with $\lambda_{b,k}^{\mathrm{MRT}}=\frac{1}{M\hat{\beta}_{b,k,b}}$. Here, $\lambda_{b,k}^{\mathrm{MRT}}$ is  chosen to satisfy the transmit power constraint at the BSs such that $\mathcal{E} \left\{\| {\bf{w}}_{b,k}^{\mathrm{MRT}}\|^2 \right\}=1$. By substituting \eqref{eq6.1} into \eqref{eq5.0}, the ECDR of user $(b,k)$ with MRT precoding becomes
\begin{align}\label{eq6.3}
\mathcal{R}_{b,k}^{\mathrm{MRT}} &=\tfrac{F}{L_d}\mathcal{E}\left\{\log_2\left(1+\gamma_{b,k}^{\mathrm{MRT}} \right) \right\},
\\
\label{eq6.4}
\gamma_{b,k}^{\mathrm{MRT}} &=\frac{P_{b,k}^{\mathrm{s,MRT}}}{\sum\nolimits_{(j,l) \in \mathbb{U}_{b,k}}\!\!P_{b,k,j,l}^{\mathrm{i,MRT}} \!+\!  \sum\nolimits_{(j,l) \in \mathbb{V}_{b,k}}\!\! P_{b,k,j,l}^{\mathrm{e,MRT}} \!+\! 1},
\end{align}
where $P_{b,k}^{\mathrm{s,MRT}}\!=\!E_{b,k} \lambda_{b,k}^{\mathrm{MRT}}| {\hat{\bf{h}}}_{b,k,b}^{\mathrm H}{\hat{\bf{h}}}_{b,k,b}|^2$ is the desired signal power, $P_{b,k,j,l}^{\mathrm{i,MRT}}\!=\!E_{j,l}\lambda_{j,l}^{\mathrm{MRT}} |{\hat{\bf{h}}}_{b,k,j}^{\mathrm H}{\hat{\bf{h}}}_{j,l,j}|^2$ is the intra- and inter-cell interference power, and $P_{b,k,j,l}^{\mathrm{e,MRT}}\!=\! E_{j,l}\lambda_{j,l}^{\mathrm{MRT}} | {\tilde{\bf{h}}}_{b,k,j}^{\mathrm H}{\hat{\bf{h}}}_{j,l,j} |^2$ is the interference power caused by the CSI estimation error.

\vspace{-0.5em}
\begin{prop} \label{prop1}
\emph{For MRT, the ECDR of user $(b,k)$ is lower bounded as follows
\begin{align}\label{eq6.6}
\mathcal{R}_{b,k}^{\mathrm{MRT}}\geq\tilde{\mathcal{R}}_{b,k}^{\mathrm{MRT}}
=\tfrac{F}{L_d}\log_2\left(1+\tilde{\gamma}_{b,k}^{\mathrm{MRT}}\right),
\end{align}
where $\tilde{\gamma}_{b,k}^{\mathrm{MRT}}$ is a lower bound for the SINR of user $(b,k)$ given as follows
\begin{equation}\label{eq6.7} 
\tilde{\gamma}_{b,k}^{\mathrm{MRT}}
\!\!=\!\!\frac{\frac{(M-1)(M-2)}{M}E_{b,k} \hat{\beta}_{b,k,b}} {\!\!\sum\limits_{\!\!(j,l)\in \mathbb{U}_{b,k,b}\!\!\!\!\!\!\!\!\!\!\!\!\!}\!\!\frac{M\!-\!2}{M}E_{b,l}\hat{\beta}_{b,k,b}
\!\!+\!\!\!\sum\limits_{(j,l) \in \mathbb{D}_1\!\!}\!\!\!\!E_{j,l}\hat{\beta}_{b,k,j}
\!+\!\!\!\sum\limits_{(j,k) \in \mathbb{D}_2\!\!}\!\!(M\!\!+\!\!1)E_{j,k}\hat{\beta}_{b,k,j}
\!\!+\!\!\!\sum\limits_{(j,l) \in \mathbb{U}_{b,k}\!\!\!\!\!\!\!\!\!\! }\!\!  E_{j,l}\tilde{\beta}_{b,k,j}
\!\!+\!\frac{M\!-\!2}{M}E_{b,k}\tilde{\beta}_{b,k,b}
\!\!+\! \!1}.\!\!
\end{equation}
Here, $\mathbb{D}_1 \stackrel{\Delta}{=} \{(j,l)\in \mathbb{U}_{b,k} \mid j\neq b,l\neq k\}$ and $\mathbb{D}_2 \stackrel{\Delta}{=} \{(j,l) \in \mathbb{U}_{b,k} \mid j\neq b, l=k\}$ define two types of inter-cell interfering users for user $(b,k)$, where a \emph{Type II} user $(j,k)\in \mathbb{D}_2$  employs the same pilot sequence as user $(b,k)$ during channel estimation, but a \emph{Type I} user $(j,l)\in \mathbb{D}_1$ does not. Moreover, $\hat{\beta}_{b,k,j}$ and $\tilde{\beta}_{b,k,j}$ are the variances of the channel estimate and the estimation error given in \eqref{eq3.5} and \eqref{eq3.6}, respectively.
}
\end{prop}
\vspace{-0.5em}
\begin{IEEEproof}
Please refer to Appendix~\ref{proof1}.
\end{IEEEproof}

\begin{rem} 
Considering \eqref{eq6.6} and \eqref{eq6.7}, caching can improve the ECDR of MRT in two respects. On the one hand, the number of interfering users in $\mathbb{U}_{b,k,j}$ is reduced by cache-enabled offloading and interference cancellation. On the other hand, the estimation accuracy of $\hat{\beta}_{b,k,j}$ is improved, i.e., the estimation error $\tilde{\beta}_{b,k,j}$ reduces due to cache-enabled offloading.
\end{rem}

\begin{rem} \label{rem2}
The interference caused by Type I and Type II users is given by the second and third terms in the denominator of \eqref{eq6.7}, respectively. Eq. \eqref{eq6.7} reveals that the interference caused by a Type II user is much larger than that caused by a Type I user when $M+1 \gg 1$, as the CSI estimates of user $(b,k)$ and Type II users are collinear in the presence of pilot contamination. Hence, reducing the number of Type II users by exploiting cache-enabled offloading at these users together with cache-enabled interference cancellation at user $(b,k)$ can significantly improve $\tilde{\gamma}_{b,k}^{\mathrm{MRT}}$ and $\mathcal{R}_{b,k}^{\mathrm{MRT}}$.
\end{rem}

\vspace{-1.5em}
\subsection{Zero-Forcing Precoding} \label{Zero-Forcing Precoding}
Different from MRT, ZF precoding is employed at the BS to mitigate multiuser interference, whereby the transmit signal of each user is projected into the null space of all other users within the \emph{same} cell \cite{Zhu2016,Chien2016}. When files are cached at the users, the interference cancellation capabilities offered by caching and ZF precoding can be combined for improved precoder design. In particular, on the one hand, if $c_{b,l,b,l}=0$, i.e., user $(b,l)$ is inactive, its channel will not be included in the design of user $(b,k)$'s ZF precoder. On the other hand, if user $(b,l)$ is active and has cached the file requested by user $(b,k)$, i.e., $c_{b,l,b,l}=1$ and $c_{b,k,b,l}=0$, user $(b,l)$ can exploit the cached file to remove the interference caused by user $(b,k)$ without relying on ZF precoding. Hence, the ZF precoder intended for user $(b,k)$ only needs to avoid causing interference to the set of active users in cell $b$ that do not have user $(b,k)$'s requested file in their caches. This set of users is denoted by $\mathbb{N}_{b,k} \triangleq \left\{ (j,l) \in\{b\}\times \mathbb{K} \mid c_{j,l,j,l}\!=\!1, c_{b,k,j,l}\!=\!1, l \neq k  \right\}$. Let $N_{b,k}^{\mathrm{n}}$ be the cardinality of $\mathbb{N}_{b,k}$, i.e., $N_{b,k}^{\mathrm{n}}=\left| \mathbb{N}_{b,k} \right|$. Consequently, if $\mathbb{N}_{b,k}$ is non-empty, the precoding vector ${\bf{w}}_{b,k}^{\mathrm{ZF}}$ of user $(b,k)$ has to satisfy the following constraints:
\begin{align}\label{eq7.1}
& \left\| {\bf{w}}_{b,k}^{\mathrm{ZF}} \right\|^2=1 \ \mathrm{and} \ {\hat{\bf{h}}}_{j,l,b}^{\mathrm H} {\bf{w}}_{b,k}^{\mathrm{ZF}}=0, \forall (j,l) \in \mathbb{N}_{b,k},
\end{align}
such that the signal of user $(b,k)$ is transmitted in the null space of the signal space spanned by the channels of the users in $\mathbb{N}_{b,k}$. Let $\mathbb{N}_{b,k} (n)$ be the $n$th element of set $\mathbb{N}_{b,k}$, where $n=1,2,\ldots,N_{b,k}^{\mathrm{n}}$. Then, for user $(b,k)$ and set $\mathbb{N}_{b,k}$, we define the effective channel estimate matrix after cache-enabled offloading and interference cancellation as ${\bf{Q}}_{b,k}=[{\bf{q}}_{b,k,1},\ldots,{\bf{q}}_{b,k,N_{b,k}^{\mathrm{n}}+1}]$, where ${\bf{q}}_{b,k,1}={\hat{\bf{h}}}_{b,k,b}$ and ${\bf{q}}_{b,k,n+1}={\hat{\bf{h}}}_{b,\mathbb{N}_{b,k} (n),b}$, $n = 1,2,\ldots,N_{b,k}^{\mathrm{n}}$. Consequently, for the proposed cache-aided massive MIMO, the ZF precoder of user $(b,k)$ exists if $M>N_{b,k}^{\mathrm{n}}+1$ and is given as follows
\begin{equation}\label{eq7.3}
{\bf{w}}_{b,k}^{\mathrm{ZF}}=\sqrt{\lambda_{b,k}^{\mathrm{ZF}}} {{\bf{Q}}_{b,k}({\bf{Q}}_{b,k}^{\rm H}{\bf{Q}}_{b,k})^{-1}{\bf{e}}_1},
\end{equation}
where ${\bf{e}}_1 \triangleq \left[1,0,\ldots,0 \right]^{\mathrm{T}}\in{\mathbb{C}^{\left(N_{b,k}^{\mathrm{n}}+1\right) \times 1}}$, $\lambda_{b,k}^{\mathrm{ZF}}$ is a normalization constant chosen to ensure $\mathcal{E} \left\{\| {\bf{w}}_{b,k}^{\mathrm{ZF}}\|^2 \right\}$ $=1$ and its value is given in Lemma \ref{lemma1}.

\vspace{-0.5em}
\begin{lem} \label{lemma1}
\emph{When $M\geq N_{b,k}^{\mathrm{n}}+1$, the normalization constant $\lambda_{b,k}^{\mathrm{ZF}}$ is given as follows
\begin{equation}\label{eq7.4}
\lambda_{b,k}^{\mathrm{ZF}}=\left(M-N_{b,k}^{\mathrm{n}}-1\right){\hat{\beta}_{b,k,b}}.
\end{equation}
}
\end{lem}
\vspace{-0.5em}
\begin{IEEEproof}
Please refer to Appendix~\ref{lemmaproof1}.
\end{IEEEproof}
\vspace{-0.5em}
\begin{prop} \label{prop2}
\emph{
For the ZF precoder ${\bf{w}}_{b,k}^{\mathrm{ZF}}$ in (\ref{eq7.3}), the ECDR 
of user $(b,k)$ is given as follows
\begin{align}\label{eq7.5}
\mathcal{R}_{b,k}^{\mathrm{ZF}} &=\tfrac{F}{L_d}\mathcal{E}\left\{\log_2\left(1+\gamma_{b,k}^{\mathrm{ZF}} \right) \right\}, \\
\label{eq7.6}
\gamma_{b,k}^{\mathrm{ZF}} &=\frac{E_{b,k} \lambda_{b,k}^{\mathrm{ZF}}} {\sum\nolimits_{(j,l) \in \mathbb{U}_{b,k} \setminus \mathbb{U}_{b,k,b} \!}\! P_{b,k,j,l}^{\mathrm{i,ZF}} \!+\! \sum\nolimits_{(j,l) \in \mathbb{V}_{b,k}\!}\! P_{b,k,j,l}^{\mathrm{e,ZF}} \!+\! 1},
\end{align}
where $P_{b,k,j,l}^{\mathrm{i,ZF}}\!=\!E_{j,l}\lambda_{j,l}^{\mathrm{ZF}}| {\hat{\bf{h}}}_{b,k,j}^{\mathrm H}{\bf{Q}}_{j,l}({\bf{Q}}_{j,l}^{\rm H}{\bf{Q}}_{j,l})^{-1}{\bf{e}}_1 |^2$ and $P_{b,k,j,l}^{\mathrm{e,ZF}}\!=\! E_{j,l}\lambda_{j,l}^{\mathrm{ZF}}| {\tilde{\bf{h}}}_{b,k,j}^{\mathrm H}{\bf{Q}}_{j,l}({\bf{Q}}_{j,l}^{\rm H}{\bf{Q}}_{j,l})^{-1}{\bf{e}}_1 |^2$ are the residual inter-cell interference power and the interference power caused by the CSI estimation errors, respectively.
Moreover, $\mathcal{R}_{b,k}^{\mathrm{ZF}}$ is lower bounded by
\begin{align}
\tilde{\mathcal{R}}_{b,k}^{\mathrm{ZF}}&=\tfrac{F}{L_d}\log_2\left(1+\tilde{\gamma}_{b,k}^{\mathrm{ZF}}\right), \label{eq7.8}\\
\tilde{\gamma}_{b,k}^{\mathrm{ZF}}
\!&=\!\frac{\left(M\!-\!N_{b,k}^{\mathrm{n}}\!-\!1\right)E_{b,k}\hat{\beta}_{b,k,b}}{\!\sum\nolimits_{(j,l)\in \mathbb{D}_3} {E_{j,l}\hat{\beta}_{b,k,j}}\!+\!\!\sum\nolimits_{j \in \mathbb{D}_2\!}\!\left(M\!-\!N_{j,k}^{\mathrm{n}}\!-\!1\right)E_{j,k}\hat{\beta}_{b,k,j}
\!+\!\sum\nolimits_{(j,l) \in \mathbb{V}_{b,k}}\! E_{j,l}\tilde{\beta}_{b,k,j}\!+\!1}, \label{eq7.9}
\end{align} 
where $\mathbb{D}_3 \triangleq \{(j,l) \in \mathbb{D}_{1} \mid (j,k)\notin \mathbb{N}_{j,l}\} \subseteq \mathbb{D}_{1} $ is the set of residual Type I inter-cell interfering users for user $(b,k)$ after employing the proposed ZF precoding\footnote{Note that, if $(j,k)\in \mathbb{N}_{j,l}$, user $(j,l)$ does not cause interference to user $(j,k)$ over channel $\hat{\bf{h}}_{j,k,j}$ due to the ZF precoding. As $\hat{\bf{h}}_{j,k,j}$ and $\hat{\bf{h}}_{b,k,b}$ are collinear, user $(j,l)$ will also not cause interference to user $(b,k)$ and, hence, is excluded from $\mathbb{D}_3$.}. }
\end{prop}
\vspace{-0.5em}
\begin{IEEEproof}
Please refer to Appendix~\ref{proof2}.
\end{IEEEproof}

\begin{rem}
Eqs. \eqref{eq7.5} and \eqref{eq7.6} show that, in contrast to MRT precoding, the intra-cell interference can be completely mitigated by the joint design of ZF precoding at the BS and caching at the users when $M> N_{b,k}^{\mathrm{n}}+1$. Consequently, due to cache-enabled interference cancellation and offloading, more spatial degrees of freedom are available for ZF precoding design. Moreover, the transmit power of each active user can be increased as $N_{b,k}^{\mathrm{n}}$ reduces due to caching. Additionally, with ZF precoding, the proposed scheme also reduces the number of interfering users in set $\mathbb{U}_{b,k}$ and the interference caused by the CSI estimation error. In particular, the inter-cell interference to user $(b,k)$ caused by Type I and Type II users is given by the first and second terms of the denominator in \eqref{eq7.9}, respectively. Similar to MRT, with ZF precoding, the interference from a Type II user dominates that from a Type I user when $M\!-\!N_{j,k}^{\mathrm{n}}\!-\!1 \gg 1$. However, the interference caused by Type II users scales with $M\!-\!N_{j,k}^{\mathrm{n}}\!-\!1$ in \eqref{eq7.9}, rather than with $M+1$ as in \eqref{eq6.7}, as a portion of the interference is eliminated by the ZF precoding.
\end{rem}

\vspace{-1em}
\subsection{Regularized Zero-Forcing Precoding} \label{Regularized Zero-Forcing Precoding}
RZF precoding is often considered in massive MIMO systems to balance interference mitigation and power enhancement \cite{Zhu2016}.
For the proposed cache-aided massive MIMO, RZF precoding has to be reconsidered in order to maximize the performance gains enabled by caching. Meanwhile, the ECDR of RZF precoding cannot be analyzed in the same manner as that of MRT and ZF precoding. To make the analysis tractable, we investigate RZF precoding in the large system limit, when $M, K \rightarrow \infty$ but $\rho_0$ is fixed.

For notational convenience, we write 
${\hat{\bf{H}}}_{b,j}$ as
\begin{align}\label{eq8.1}
{\hat{\bf{H}}}_{b,j} = \sqrt{M} {\hat{\bf{D}}}_{b,j}^{\frac{1}{2}} {\hat{\bf{G}}}_{b,j}, 
\end{align}
with ${\hat{\bf{D}}}_{b,j}=\mathrm{diag}([\hat{\beta}_{{b,1,j}},\ldots,\hat{\beta}_{{b,K,j}}])\in\!{{\mathbb C}^{{K} \times {K}}}$ and ${\hat{\bf{G}}}_{b,j}=[{\hat{\bf{g}}}_{b,1,j},\ldots,{\hat{\bf{g}}}_{b,K,j}]^{\mathrm{T}}\in\!{{\mathbb C}^{{K} \times {M}}}$.
The elements of ${\hat{\bf{G}}}_{b,j}$ are independently and identically distributed as $\mathcal{CN}(0,1/M)$\cite{Zhu2016,Sifaou2014}. For user $(b,k)$, we define the effective channel fading estimation matrix after cache-enabled offloading and interference cancellation as ${\bf{F}}_{b,k}=[{\bf{f}}_{b,k,1},\ldots,{\bf{f}}_{b,k,N_{b,k}^{\mathrm{n}}+1}]^{\mathrm{T}}$, where ${\bf{f}}_{b,k,1}={\hat{\bf{g}}}_{b,k,b}$, and ${\bf{f}}_{b,k,n+1}={\hat{\bf{g}}}_{b,\mathbb{N}_{b,k} (n),b}$, $n = 1,2,\ldots,N_{b,k}^{\mathrm{n}}$. Then, the RZF precoding vector of user $(b,k)$ is given as  \cite{Nguyen2008, Zhu2016}
\begin{equation}\label{eq8.3}
{\bf{w}}_{b,k}^{\mathrm{RZF}}=\sqrt{\lambda_{b,k}^{\mathrm{RZF}}} ({\bf{F}}_{b,k}^{\rm H}{\bf{F}}_{b,k}+\alpha_{b,k}{\bf{I}}_{M})^{-1}{\bf{f}}_{b,k,1},
\end{equation}
where $\alpha_{b,k}$ is the regularization parameter that can be adjusted to further increase the achievable rate and $\lambda_{b,k}^{\mathrm{RZF}}$ is a normalization constant chosen to ensure $\mathcal{E} \left\{\| {\bf{w}}_{b,k}^{\mathrm{RZF}}\|^2 \right\}=1$. When the $\beta_{b,k,b}$s in cell $b$ are equal, the RZF precoder in \eqref{eq8.3} converges to the ZF precoder in \eqref{eq7.3} as $\alpha_{b,k}\rightarrow 0$ and the MRT precoder in \eqref{eq6.1} as $\alpha_{b,k}\rightarrow \infty$.

\vspace{-0.5em}
\begin{lem} \label{lemma2}
\emph{When $M, K \rightarrow \infty$ but $\rho_0$ is finite and fixed, $\lambda_{b,k}^{\mathrm{RZF}}$ is given as follows
\begin{equation}\label{eq8.4}
\lambda_{b,k}^{\mathrm{RZF}}={\left(1+\mathcal{G}_{b,k}\right)^2} / {\overline{\mathcal{G}}_{b,k}},
\end{equation}
where $\mathcal{G}_{b,k} \triangleq \mathcal{G}(\rho^{-1}_{b,k},\alpha_{b,k})$, $\overline{\mathcal{G}}_{b,k} \triangleq {-\frac{d}{d\alpha_{b,k}}\mathcal{G}(\rho^{-1}_{b,k},\alpha_{b,k})}$, $\rho_{b,k}=M/N_{b,k}^{\mathrm{n}}$, and $\mathcal{G}(\rho^{-1}_{b,k},\alpha_{b,k})$ can be evaluated in closed form as \cite{Nguyen2008}
\begin{equation}
 \mathcal{G}(\rho^{-1}_{b,k},\alpha_{b,k})\!=\!\frac{1}{2}\left[ \sqrt{\frac{(1-\rho^{-1}_{b,k})^2}{\alpha_{b,k}^2}+\frac{2(1+\rho^{-1}_{b,k})}{\alpha_{b,k}}+1} + \frac{1-\rho^{-1}_{b,k}}{\alpha_{b,k}} \!-1\! \right].
\end{equation}
}
\end{lem}
\vspace{-0.5em}
\begin{IEEEproof}
Please refer to Appendix~\ref{lemmaproof2}.
\end{IEEEproof}

By substituting \eqref{eq8.3} into \eqref{eq5.1} and \eqref{eq5.2}, the ECDR of user $(b,k)$ with RZF precoding is given as follows
\begin{align}\label{eq8.6}
\mathcal{R}_{b,k}^{\mathrm{RZF}} &=\tfrac{F}{L_d}\mathcal{E}\left\{\log_2\left(1+\gamma_{b,k}^{\mathrm{RZF}} \right) \right\}, \\
\label{eq8.7}
\gamma_{b,k}^{\mathrm{RZF}} &=\frac{P_{b,k}^{\mathrm{s,RZF}}}{\sum\nolimits_{(j,l) \in \mathbb{U}_{b,k}\!}\! P_{b,k,j,l}^{\mathrm{i,RZF}} \!+\! \sum\nolimits_{(j,l) \in \mathbb{V}_{b,k}\!}\! P_{b,k,j,l}^{\mathrm{e,RZF}} \!+\! 1},
\end{align}
where $\!P_{b,k}^{\mathrm{s,RZF}}\!\!\!=\!E_{b,k} \lambda_{b,k}^{\mathrm{RZF}}| {\hat{\bf{h}}}_{b,k,b}^{\mathrm H}({\bf{F}}_{\!b,k}^{\rm H}{\bf{F}}_{\!b,k}\!+\!\alpha_{b,k}{\bf{I}}_{M})^{\!-\!1\!}{\bf{f}}_{b,k,1}|^2$, $P_{b,k,j,l}^{\mathrm{i,RZF}}\!\!=\!E_{j,l}\lambda_{j,l}^{\mathrm{RZF}}| {\hat{\bf{h}}}_{b,k,j}^{\mathrm H}({\bf{F}}_{\!j,l}^{\rm H}{\bf{F}}_{\!j,l}\!+\!\alpha_{j,l}{\bf{I}}_{M})^{\!-\!1\!}{\bf{f}}_{j,l,1} |^2\!$, and $P_{b,k,j,l}^{\mathrm{e,RZF}}\!\!=\!E_{j,l}\lambda_{j,l}^{\mathrm{RZF}}| {\tilde{\bf{h}}}_{b,k,j}^{\mathrm H}({\bf{F}}_{\!j,l}^{\rm H}{\bf{F}}_{\!j,l}\!+\!\alpha_{j,l}{\bf{I}}_{M})^{\!-\!1\!}{\bf{f}}_{j,l,1} |^2$.

\vspace{-0.5em}
\begin{prop} \label{prop3}
\emph{
When $M, K \rightarrow \infty$ but $\rho_0$ is finite and fixed, for the RZF precoder ${\bf{w}}_{b,k}^{\mathrm{RZF}}$ in (\ref{eq8.3}), the ECDR of user $(b,k)$, $\mathcal{R}_{b,k}^{\mathrm{RZF}}$, is given as follows
\vspace{-0.5em}
\begin{align}\label{eq8.9}
\mathcal{R}_{b,k}^{\mathrm{RZF}} &=\tfrac{F}{L_d}\log_2\left(1+{\gamma}_{b,k}^{\mathrm{RZF}}\right), \\
\label{eq8.10} 
{\gamma}_{b,k}^{\mathrm{RZF}}
 &= \tfrac{{ E_{b,k} \hat{\beta}_{b,k,b} \mathcal{G}^2_{b,k}} / {\overline{\mathcal{G}}_{b,k}}}
{ \sum\limits_{(j,l)\in \mathbb{U}_{b,k,b}} \!\!\!\!\!\! \frac{E_{b,l}\hat{\beta}_{b,k,b}}{M \left(1+\mathcal{G}_{b,l}\right)^2} + \!\!\! \sum\limits_{(j,l)\in \mathbb{D}_3} \!\!\! \frac{E_{j,l}\hat{\beta}_{b,k,j}}{M} + \!\!\! \sum\limits_{(j,l)\in \mathbb{D}_4} \!\!\! \frac{E_{j,l}\hat{\beta}_{b,k,j}} {M(1+\mathcal{G}_{j,l})^2} + \!\!\! \sum\limits_{j \in
\mathbb{D}_2} \!\!\! \frac{ E_{j,k}\hat{\beta}_{b,k,j} \mathcal{G}^2_{j,k}}{\overline{\mathcal{G}}_{j,k} } + \!\!\!\!\! \sum\limits_{(j,l) \in \mathbb{V}_{b,k} } \!\!\!\!\! \frac{E_{j,l}{\tilde{\beta}_{b,k,j}}}{M} +\frac{1}{M}},
\end{align}
where $\mathbb{D}_4 \triangleq \{(j,l) \in \mathbb{D}_{1} \mid (j,k)\in \mathbb{N}_{j,l} \} \subseteq \mathbb{D}_{1}$ is the set of residual Type I inter-cell interfering users for user $(b,k)$ after employing the proposed RZF precoding. Note that $\mathbb{D}_3$ and $\mathbb{D}_4$ constitute a partition of $\mathbb{D}_1$, i.e., $\mathbb{D}_3 \cap \mathbb{D}_4 =\emptyset$ and $\mathbb{D}_3 \cup \mathbb{D}_4 = \mathbb{D}_1$.
}
\end{prop}
\vspace{-0.5em}
\begin{IEEEproof}
Please refer to Appendix~\ref{proof3}.
\end{IEEEproof}

\begin{rem}
{{In \eqref{eq8.10}, the sizes of sets $\mathbb{U}_{b,k,b}$, $\mathbb{D}_3$, $\mathbb{D}_4 $, and $\mathbb{V}_{b,k}$ can scale with $M$, whereas the cardinality of set $\mathbb{D}_2$ is independent of $M$. Hence, \eqref{eq8.10} has to be understood as the limit of ${\gamma}_{b,k}^{\mathrm{RZF}}$, if exists, when $M, K \rightarrow \infty$ with $\rho_0$ being finite and fixed.}} 
\end{rem}

\begin{rem}
Proposition~\ref{prop3} shows that, in contrast to MRT and ZF precoding, joint caching and RZF precoding impacts not only the transmit power, cf. the number of users interfering the user of interest, $N_{b,k}^{\mathrm{n}}$, but also the desired signal and the interference powers, cf. function $\mathcal{G}$ and user set $\mathbb{U}_{b,k}$.
Hence, a tradeoff between the signal power and the interference power, which is adjusted by both the regularization parameter ${\alpha_{b,k}}$ and the cache status, exists and will be investigated numerically in Section~V for maximization of the cache-enabled performance gains in massive MIMO systems with RZF precoding.
\end{rem}

\vspace{-1em}
\section{Performance Analysis of Joint Caching and Precoding} \label{Cache Placement Scheme}
In Section~\ref{Cache-Aided Linear Precoder Design}, the performance of the proposed precoding schemes has been analyzed for an arbitrary given cache status of the users. In this section, we evaluate the impact of joint cache placement and precoding on the performance of cache-aided massive MIMO. We consider two popular cache placement schemes, namely (random) uncoded caching \cite{Wei2019,Malak2014,Blaszczyszyn2015} and coded caching \cite{Maddah-Ali2014, Ngo2018, TWC22a, TWC22b}. {{For random uncoded caching, each user caches an entire file with a given probability. Deterministic uncoded caching is only a special case of the considered random uncoded caching. For coded caching, the users cache portions of each file in a coded format.}}

We analyze and compare the ECDRs achieved by uncoded and coded caching in combination with the proposed precoding schemes.
Throughout this section, we assume that the users in cell $j\in \mathbb{B}$ request file $l_s\in \mathbb{L}_s$ with probability $q_{j,l_s}^{\mathrm r}$, where $\sum\nolimits_{l_s \in \mathbb{L}_s} q_{j,l_s}^{\mathrm r}=1$. For example, a Zipf distribution based file popularity model has been adopted in the literature \cite{Malak2014,Blaszczyszyn2015}. In this case, $q_{j,l_s}^{\mathrm r}$ is given as follows
\begin{equation}\label{eq9.1}
q_{j,l_s}^{\mathrm r}=\frac{{1}/{l_s^{\eta_j}}}{\sum\nolimits_{l_i \in \mathbb{L}_s} {1}/{l_i^{\eta_j}}},
j\in \mathbb{B},l_s\in \mathbb{L}_s,
\end{equation}
where $\eta_j$ is the Zipf exponent of cell $j$ and $\eta_j \in (0,1), \forall j$. A large $\eta_j$ indicates a high correlation among the requested files for the users in cell $j$ \cite{Malak2014,Blaszczyszyn2015}. Moreover, to gain insight, we assume uniform transmit power allocation to the active users, i.e., $E_{j,l}=E_0/\overline{K}_j$.

\vspace{-1em}
\subsection{(Random) Uncoded Caching} \label{Binary Caching Scheme}
Assume that each user in cell $j\in \mathbb{B}$ caches file $l_s\in \mathbb{L}_s$ with probability $q_{j,l_s}^{\mathrm c} \in [0,1]$. According to \cite{Malak2014,Blaszczyszyn2015}, the placement probabilities are typically selected such that the files cached at the users do not exceed the cache size on average, i.e., $\sum\nolimits_{l_s \in \mathbb{L}_s} q_{j,l_s}^{\mathrm c}\leq L_u,\forall j$. For example, for $q_{j,l_s}^{\mathrm c} = L_u / L_s$ all files are cached with the same probability.

Note that we have assumed all users within a cell to have the same request and placement probabilities. Hence, for all users in cell $j\in \mathbb{B}$, the probability that a given user is active is denoted by $q_{j}^{\mathrm{a}}$ for uncoded caching. Similarly, we denote the probability that a user in cell $j$ belongs to set $\mathbb{N}_{j,l}$ by $q_{j}^{\mathrm{n}}$ (recall that $\mathbb{N}_{j,l}$ is the set of active users in cell $j$ that do not have user $(j,l)$'s requested file in their caches), and the probability that the file requested by an active user in cell $j'$ is not cached at any user in cell $j$ by $q_{j,j'}^{\mathrm{i}}$. Then, for the considered (random) uncoded caching with request probabilities $\{q_{j,l_s}^{\mathrm{r}}\}$ and placement probabilities $\{q_{j,l_s}^{\mathrm{c}}\}$, it follows from the total law of probability \cite{Kokoska2000} that
\vspace{-0.5em}
\begin{subequations} \label{eq9.3}
\begin{align}
\label{eq9.3a}q_j^{\mathrm{a}}&=\sum\nolimits_{l_s \in \mathbb{L}_s} q_{j,l_s}^{\mathrm r}\left(1\!-\!q_{j,l_s}^{\mathrm c}\right),  \\
\label{eq9.3b}q_{j,j'}^{\mathrm{i}}&=\sum\nolimits_{l_s \in \mathbb{L}_s} q_{j',l_s}^{\mathrm r}\left(1\!-\!q_{j',l_s}^{\mathrm c}\right)\!\left(\sum\nolimits_{l_i\neq l_s} q_{j,l_i}^{\mathrm r} \left(1\!-\!q_{j,l_i}^{\mathrm c}\right)\!\left(1\!-\!q_{j,l_s}^{\mathrm c}\right)\!+\!q_{j,l_s}^{\mathrm r} \left(1\!-\!q_{j,l_s}^{\mathrm c}\right)\right),\\
\label{eq9.3c}q_{j}^{\mathrm{n}}&=\sum\nolimits_{l_s\in \mathbb{L}_s} q_{j,l_s}^{\mathrm r}\left(1\!-\!q_{j,l_s}^{\mathrm c}\right)^2\left(\sum\nolimits_{l_i\neq l_s} q_{j,l_i}^{\mathrm r} \left(1\!-\!q_{j,l_i}^{\mathrm c}\right)\!+\!q_{j,l_s}^{\mathrm r}\right).
\end{align}
\end{subequations}

Moreover, the ECDR defined in \eqref{eq5.0} assumes the file delivery size $L_d$ or the delivery time $L_d / R_{b,k}$ to be deterministic. However, with random uncoded caching, $L_d$ and $L_d / R_{b,k}$ become random variables. To account for the randomness, we generalize the definition of the ECDR by replacing $L_d / R_{b,k}$  in \eqref{eq5.0} with the {expected} delivery time, namely $(1- q_{b}^{\mathrm{a}}) \times 0  + q_{b}^{\mathrm{a}} \times F/R_{b,k} = q_{b}^{\mathrm{a}} F/R_{b,k} $. The resulting ECDR, referred to as the \emph{average} ECDR, of random uncoded caching is given as 
\begin{equation} \label{eq5.0.2}
\mathcal{R}_{b,k}^{\mathrm{u}}=R_{b,k} / q_{b}^{\mathrm{a}}.
\end{equation} 
In \eqref{eq5.0.2}, the offloading gains enabled by random uncoded caching lead to a reduction in the expected delivery time by $q_{b}^{\mathrm{a}}$  and hence, a scaling of the average ECDR by factor $1/ q_{b}^{\mathrm{a}} \ge 1$. Note that $\mathcal{R}_{b,k}^{\mathrm{u}}$ in \eqref{eq5.0.2} reduces to $\mathcal{R}_{b,k}$ in \eqref{eq5.0} when conditioning on user $(b,k)$ to be active or inactive. In the following, focusing on the user $(b,k)$ of interest, we analyze the joint impact of $q_b^{\mathrm{a}}$, $q_{b,j}^{\mathrm{i}}$, and $q_{b}^{\mathrm{n}}$ for random uncoded caching in combination with the proposed MRT, ZF, and RZF precoding schemes. For comparison, \emph{conventional} massive MIMO without (uncoded) caching is also analyzed as a baseline, and is henceforth referred to as Baseline Scheme~$1$. {{Unlike Section~\ref{Cache-Aided Linear Precoder Design}, here we eliminate the assumption that user $(b,k)$ is active.}}

\subsubsection{MRT} \label{MRT}
Due to the randomness of the users' cached and requested files, sets $\mathbb{U}_{b,k,j}$ and $\mathbb{U}_{b,k}$ in \eqref{eq6.7} are random. To facilitate the analysis, we assume $M$, $K$ to be asymptotically large, i.e., $M, K\rightarrow \infty$, while keeping the ratio $\rho_0 = {M}/{K}$ to be fixed.
\vspace{-0.5em}
\begin{lem} \label{lem5}
\emph{
With the considered uncoded caching scheme and uniform power allocation, a lower bound for the ECDR achieved at user $(b,k)$ when $M, K\rightarrow \infty$ with $\rho_0$ being fixed is given as follows 
\vspace{-0.5em}
\begin{align}\label{eq10.1}
{{
\tilde{\mathcal{R}}_{b,k}^{\mathrm{MRT,P1}} \!\!=\!\! \frac{1}{q_{b}^{\mathrm{a}}}\log_2 \!\left(\! 1 \!+\! \frac{{\rho_0}\hat{\beta}_{b,k,b} / {q_b^{\mathrm{a}}}} {{q_{b,b}^{\mathrm{i}} \hat{\beta}_{b,k,b}}/{q_b^{\mathrm{a}}}
+ \sum\nolimits_{j \neq b}  {q_{b,j}^{\mathrm{i}}(\rho_0+1)\hat{\beta}_{b,k,j}}/{q_j^{\mathrm{a}}}
+ \sum\nolimits_{j \in \mathbb{B}} {q_{b,j}^{\mathrm{i}}\tilde{\beta}_{b,k,j}} / {q_{j}^{\mathrm{a}}}+ {1}\!/\!{E_0}} \!\right).}}
\end{align} 
}
\end{lem}  
\vspace{-0.5em}
\begin{IEEEproof}
The result follows from Proposition~\ref{prop1}. Particularly, by the law of large numbers (LLN) \cite{Smythe1973}, when $K\rightarrow \infty$, the number of active users in cell $j\in \mathbb{B}$ converges to its mean value, $q_j^{\mathrm{a}}K$, with probability one. Hence, with uniform power allocation the transmit power $E_{j,l}$ converges to $E_0/(q_j^{\mathrm{a}}K)$. Moreover, the number of users in set $\mathbb{U}_{b,k,j}$ converges to its mean value, $q_{b,j}^{\mathrm{i}}K$. In this case, the first term in the denominator of \eqref{eq6.7} converges to $\frac{q_{b,b}^{\mathrm{i}} \hat{\beta}_{b,k,b} E_0}{q_b^{\mathrm{a}}}$. Likewise, the other terms in the denominator of \eqref{eq6.7} converge to and, hence, can be substituted with their mean values. Finally, \eqref{eq10.1} can be obtained by letting $M, K\rightarrow \infty$ with $\rho_0$ being fixed.
\end{IEEEproof}

By setting $q_j^{\mathrm{a}}=1$ and $q_{b,j}^{\mathrm{i}}=1,j\in \mathbb{B}$ in \eqref{eq10.1}, a lower bound for the ECDR 
of Baseline Scheme~$1$ is given as follows
\begin{equation}\label{eq10.2}
\tilde{\mathcal{R}}_{b,k}^{\mathrm{MRT,B1}}=\log_2\!\left(1+ \frac{{\rho_0}\hat{\beta}_{b,k,b}^{\mathrm{b}}} {\hat{\beta}_{b,k,b}^{\mathrm{b}}
+ \sum\nolimits_{j \neq b}\! (\rho_0+1)\hat{\beta}_{b,k,j}^{\mathrm{b}}
+ \sum\nolimits_{j \in \mathbb{B}} \!\tilde{\beta}_{b,k,j}+ {1}/{E_0}} \!\right),
\end{equation}
where $\hat{\beta}_{b,k,j}^{\mathrm{b}}=\frac{p\tau\beta_{b,k,j}^2}{1+p\tau\sum\nolimits_{u \in \mathbb{B}} \beta_{u,k,j}}$ and $\tilde{\beta}_{b,k,j}^{\mathrm{b}}=\frac{(1+p\tau\sum\nolimits_{u \neq b} \beta_{u,k,j})\beta_{b,k,j}}{1+p\tau\sum\nolimits_{u \in \mathbb{B}}\beta_{u,k,j}}$.

\begin{rem}
As expected, $\tilde{\mathcal{R}}_{b,k}^{\mathrm{MRT,P1}}$ decreases monotonically with increasing $q_b^{\mathrm{a}}$ and $q_{b,j}^{\mathrm{i}}$. However, $\tilde{\mathcal{R}}_{b,k}^{\mathrm{MRT,P1}}$ increases with increasing $q_j^{\mathrm{a}},j\neq b$, as the interference power from the active users in other cells decreases with increasing $q_j^{\mathrm{a}},j\neq b$. Hence, comparing \eqref{eq10.1} and \eqref{eq10.2}, we have $\tilde{R}_{b,k}^{\mathrm{MRT,P1}} \geq \tilde{R}_{b,k}^{\mathrm{MRT,B1}}$. The performance gains of the proposed scheme over Baseline Scheme~$1$ include: i) a scaling pre-log factor $1/ q_{b}^{\mathrm{a}} \ge 1$, cf. \eqref{eq5.0.2}, and an increased transmit power per active user given by $E_0 / (q_b^{\mathrm{a}}K) > E_0/K$, both due to cache-enabled offloading, ii) reduced intra- and inter-cell interference due to cache-enabled interference cancellation  and offloading such that $q_{b,j}^{\mathrm{i}}<1$, and iii) reduced impact of CSI estimation error $\tilde{\beta}_{b,k,j}^{\mathrm{b}}$ due to cache-enabled offloading during uplink channel estimation and cache-enabled interference cancellation during data transmission such that $q_{b,j}^{\mathrm{i}}/q_j^{\mathrm{a}} \leq 1$ in the third term in the denominator of \eqref{eq10.1}. The latter inequality is due to $q_{b,j}^{\mathrm{i}}\!=\!\sum\nolimits_{l_s \in \mathbb{L}_s} \!q_{j,l_s}^{\mathrm r}\!\left(\!1\!-\!q_{j,l_s}^{\mathrm c}\!\right)\!\left(\!\sum\nolimits_{l_i\neq l_s} \!q_{b,l_i}^{\mathrm r} \!\left(\!1\!-\!q_{b,l_i}^{\mathrm c}\!\right)\!\left(\!1\!-\!q_{b,l_s}^{\mathrm c}\!\right)\!+\!q_{b,l_s}^{\mathrm r} \!\left(\!1\!-\!q_{b,l_s}^{\mathrm c}\!\right)\!\right)\!\leq \! \sum\nolimits_{l_s\in \mathbb{L}_s}\!q_{j,l_s}^{\mathrm r}\left(\!1\!-\!q_{j,l_s}^{\mathrm c}\!\right)\!\left(\sum\nolimits_{l_i \in \mathbb{L}_s} \!q_{b,l_i}^{\mathrm r} \!\right)\!=\!q_{j}^{\mathrm{a}}$, since $1\!-\!q_{b,l_i}^{\mathrm c}\leq 1$ and $1\!-\!q_{b,l_s}^{\mathrm c}\leq 1$.
\end{rem}

\subsubsection{ZF Precoding} \label{ZF}
For joint uncoded caching and ZF precoding, a lower bound on the ECDR of user $(b,k)$ when $M, K\rightarrow \infty$ with $\rho_0$ being fixed is given as follows
\begin{equation}\label{eq10.3}
{{\tilde{\mathcal{R}}_{b,k}^{\mathrm{ZF,P1}}=\! \frac{1}{q_{b}^{\mathrm{a}}} \log_2\!\left(1+\frac{{\left(\rho_0-q_{b}^{\mathrm{n}}\right)\hat{\beta}_{b,k,b}}/{q_{b}^{\mathrm{a}}}}
{\sum\nolimits_{j\neq b}{q_{b,j}^{\mathrm{i}}\left(1-2q_{j}^{\mathrm{n}}+\rho_0\right)\hat{\beta}_{b,k,j}}/{q_{j}^{\mathrm{a}}}
+\sum\nolimits_{j \in \mathbb{B}} {q_{b,j}^{\mathrm{i}}\tilde{\beta}_{b,k,j}}/{q_{j}^{\mathrm{a}}}+{1}/{E_0}}\!\right), }}
\end{equation}
which follows from \eqref{eq7.8} and the LLN, similar to \eqref{eq10.1}. By setting $q_j^{\mathrm{a}}=1$, $q_{b,j}^{\mathrm{i}}=1$, and $q_{j}^{\mathrm{n}}=1$,$j\in\{1,\ldots,B\}$ in \eqref{eq10.3}, the ECDR of user $(b,k)$ for Baseline Scheme~$1$ is lower bounded as follows
\begin{equation}\label{eq10.4}
\tilde{\mathcal{R}}_{b,k}^{\mathrm{ZF,B1}}=\log_2\left(1+\frac{(\rho_0-1)\hat{\beta}_{b,k,b}^{\mathrm{b}}}
{\sum\nolimits_{j\neq b}(\rho_0-1)\hat{\beta}_{b,k,j}^{\mathrm{b}}
+\sum\nolimits_{j \in \mathbb{B}} \tilde{\beta}_{b,k,j}^{\mathrm{b}}+{1}/{E_0}}\right).
\end{equation}

\begin{rem}
Unlike $\tilde{R}_{b,k}^{\mathrm{MRT,P1}}$, $\tilde{R}_{b,k}^{\mathrm{ZF,P1}}$ decreases not only when $q_b^{\mathrm{a}}$ and $q_{b,j}^{\mathrm{i}}, j\neq b$, increase, but also when the probability that an active user is interfered by user $(b,k)$, i.e., $q_{j}^{\mathrm{n}}$, increases. Hence, comparing \eqref{eq10.3} and \eqref{eq10.4}, we have $\tilde{R}_{b,k}^{\mathrm{ZF,P1}} \geq \tilde{R}_{b,k}^{\mathrm{ZF,B1}}$. {{This result is expected as, in addition to the benificial effect of the pre-log scaling factor $1/ q_{b}^{\mathrm{a}} \ge 1$, the proposed scheme also allows to allocate more power to each active user as the number of active users is reduced from $K$ to $q_b^{\mathrm{a}}K$ for $q_b^{\mathrm{a}}<1$.}} Moreover, thanks to cache-enabled offloading and interference cancellation, more spatial degrees of freedom are available for the ZF precoding design as the number of ZF precoding constraints in \eqref{eq7.1} reduces from $K-1$ to $q_{b}^{\mathrm{n}}(K-1)$. Furthermore, the proposed scheme can also reduce the interference caused by the CSI estimation error due to offloading as it is taken into account by the factor $q_{b,j}^{\mathrm{i}}/q_j^{\mathrm{a}} \leq 1$.
\end{rem}

\subsubsection{RZF Precoding} \label{RZF}
Following the same approach as in \cite{Nguyen2008,Zhu2016}, we adopt the same regularization parameter for all users, i.e., $\alpha_{b,k}=\alpha$. As $K\rightarrow \infty$, the numbers of the active users in cell $b$, the users in set $\mathbb{U}_{b,k,j}$, and the users in set $\mathbb{N}_{k,j}$ in \eqref{eq8.10} converge to $q_b^{\mathrm{a}}K$, $q_{b,j}^{\mathrm{i}}K$ and $q_b^{\mathrm{n}}K$ with probability one, respectively. Thereby, for joint uncoded caching and RZF precoding, the ECDR of user $(b,k)$ can be obtained from \eqref{eq8.9} and \eqref{eq8.10}, similar to \eqref{eq10.1}, and is given as follows
\begin{equation}\label{eq10.5}
\mathcal{R}_{b,k}^{\mathrm{RZF,P1}}
\!=\! \frac{1}{q_{b}^{\mathrm{a}}}\log_2\!\left(\!1+\!\frac{{\rho_0 \hat{\beta}_{b,k,b} \mathcal{G}^2_{b} } / {(q_{b}^{\mathrm{a}} \overline{\mathcal{G}}_{b}) }}{\! \frac{q_{b,b}^{\mathrm{i}}\hat{\beta}_{b,k,b}}{q_{b}^{\mathrm{a}}\left(1+\mathcal{G}_{b}\right)^2}\!+\! \sum\nolimits_{j\neq b}\!\frac{q_{b,j}^{\mathrm{i}}\hat{\beta}_{b,k,j}}{q_{j}^{\mathrm{a}}}\!\left(\!\frac{\rho_0\mathcal{G}^2_{j} }{ \overline{\mathcal{G}}_{j} }
\!+\!\frac{q_{j}^{\mathrm{n}}}{\left(1\!+\!\mathcal{G}_{j} \right)^2}\!+1\!-\!q_{j}^{\mathrm{n}}\!\right) \!+\!\sum\nolimits_{j \in \mathbb{B}} \! \frac{q_{b,j}^{\mathrm{i}}}{q_{j}^{\mathrm{a}}}\tilde{\beta}_{b,k,j}\!+\!\frac{1}{E_0}}\!\right), 
\end{equation}
when $M, K \rightarrow \infty$ but $\rho_0$ is fixed, where $\mathcal{G}_b \triangleq \mathcal{G}(\rho^{-1}_{b},\alpha)$, $\overline{\mathcal{G}}_{b} \triangleq - \frac{d}{d\alpha}\mathcal{G}(\rho^{-1}_{b},\alpha)$, and $\rho_b = M/(q_{b}^{\mathrm{n}}K)$. Moreover, for Baseline Scheme~$1$, the ECDR of user $(b,k)$ is given as follows
\begin{equation}\label{eq10.6}
\mathcal{R}_{b,k}^{\mathrm{RZF,B1}}\!=\!\log_2\!\left(\!1\!+\!\frac{{\rho_{0} \hat{\beta}_{b,k,b}^{\mathrm{b}} \mathcal{G}^2_0  } / { \overline{\mathcal{G}}_{0} }}{\frac{\hat{\beta}_{b,k,b}^{\mathrm{b}}}{\left(1+\mathcal{G}_0 \right)^2}\!+\! \sum\nolimits_{j\neq b}\!\hat{\beta}_{b,k,j}^{\mathrm{b}} \left(\frac{\rho_{0}\mathcal{G}^2_0  }{\overline{\mathcal{G}}_{0}}
\!+\!\frac{1}{\left(1\!+\!\mathcal{G}_0 \right)^2}\!\right) \!+\!\sum\nolimits_{j \in \mathbb{B}} \tilde{\beta}_{b,k,j}^{\mathrm{b}}\!+\!\frac{1}{E_0}}\right),
\end{equation}
where $\mathcal{G}_0 \triangleq \mathcal{G}(\rho^{-1}_{0},\alpha)$ and $\overline{\mathcal{G}}_{0} \triangleq - \frac{d}{d\alpha}\mathcal{G}(\rho^{-1}_{0},\alpha)$.
\begin{rem}
Comparing $R_{b,k}^{\mathrm{RZF,P1}}$ with $R_{b,k}^{\mathrm{RZF,B1}}$, we observe that, for RZF precoding, not only the ECDR of uncoded caching increases by a pre-log scaling factor $1/ q_{b}^{\mathrm{a}} \ge 1$, but also the transmit power per active user increases as the number of active users reduces from $K$ to $q_{b}^{\mathrm{a}}K$. Moreover, with the proposed scheme, both the intra- and inter-cell interference can be mitigated exploiting cache-enabled interference cancellation and offloading such that $q_{b,j}^{\mathrm{i}}<1$ and $q_{j}^{\mathrm{n}}<1$. Furthermore, the impairment of the CSI error on the proposed scheme is reduced by a factor $q_{b,j}^{\mathrm{i}}/q_j^{\mathrm{a}} < 1$, thanks to cache-enabled offloading. Thus, we have $R_{b,k}^{\mathrm{RZF,P1}} \geq R_{b,k}^{\mathrm{RZF,B1}}$. We note that, different from MRT and ZF precoding, for RZF precoding, $R_{b,k}^{\mathrm{RZF,P1}}$ jointly depends not only on $q_b^{\mathrm{a}}$, $q_{b,j}^{\mathrm{i}}$, and $q_{j}^{\mathrm{n}}$, but also on regularization parameter $\alpha$, which can be further optimized to enlarge the performance gains over $R_{b,k}^{\mathrm{RZF,B1}}$.
\end{rem}
\vspace{-.5em}
\begin{rem}
Finding the optimal random uncoded caching policy for maximization of the ECDR is prohibitive, since the pre-log scaling factor, the transmit powers, and the interference powers jointly depend on the caching probabilities. However, in order to maximize the pre-log scaling factor of the ECDR for the users in cell $b$, it is optimal to select the $L_u$ most popular files to be cached according to the following deterministic caching policy 
\vspace{-.5em}
\begin{equation} \label{eq46_opt}
q_{b,l_s}^{\mathrm{c}} = 
\begin{cases}
1, & {\text{if } l_s=1,\ldots,L_u}, \\
0, & {\text{otherwise}},
\end{cases}
\end{equation} 
as this policy will minimize probability $q_{b}^{\mathrm{a}}$. The resulting maximum pre-log scaling factor is $1/q_{b}^{\mathrm{a},*} = 1/\sum\nolimits_{l_s = L_u +1}^{L_s} q_{j,l_s}^{\mathrm r}$.
\end{rem}
\vspace{-1.5em}
\subsection{Coded Caching} \label{Binary Caching Scheme}
Now, we investigate the joint design of the coded caching scheme proposed in \cite{Maddah-Ali2014} and the precoding schemes proposed in Section~\ref{Cache-Aided Linear Precoder Design}. For ease of presentation, we assume discrete-valued cache sizes $L_u\in\{L_s/K,2L_s/K,\ldots,(K-1)L_s/K\}$ as in \cite{Maddah-Ali2014} such that $t=\frac{L_u K}{L_s}$ is an integer. The results derived in this section can be extended to continuous-valued cache sizes by following the ``time-sharing" approach in \cite{Maddah-Ali2014}, which is omitted here due to space constraints.

With coded caching, each file is split into $C_{K}^{t}$ subfiles of size ${F} / {C_{K}^{t}}$~MBytes. Let $\mathbb{T}_w$ be any subset of $\mathbb{K}$ with cardinality $t$. As $C_K^t$ choices of $\mathbb{T}_w$ are possible, we can use $\mathbb{T}_w$ to index the subfiles. Thereby, the subfiles of file $l_s$ are denoted as $W_{l_s,\mathbb{T}_w}$ for $\mathbb{T}_w \subseteq \mathbb{K}$ with $|\mathbb{T}_w| = t$. During cache placement, each subfile is cached by $t$ out of $K$ users in each cell. To diversify the subfiles cached at the users within each cell, user $(j,l)$ caches subfile $W_{l_s,\mathbb{T}_w}$ only if $l \in \mathbb{T}_w $. Thereby, with coded caching, $C_{K-1}^{t-1}=\frac{L_u}{L_s}C_{K}^{t}$ subfiles of each file will be cached at each user. During delivery, a user requests an arbitrary file and the BS associated with the user only needs to deliver the remaining $C_{K}^{t}-C_{K-1}^{t-1}=(1-\frac{L_u}{L_s})C_{K}^{t}$ subfiles of each requested file. Hence, we have $F/L_d=L_s/(L_s-L_u)$, and
\begin{equation}\label{eq11.0}
\mathcal{R}_{b,k}=\tfrac{L_s}{\!L_{\!s}\!-\!L_{\!u}}R_{b,k}.
\end{equation}
In \eqref{eq11.0}, the offloading gains enabled by coded caching result in a reduced message length and, consequently, a scaling of the ECDR $\mathcal{R}_{b,k}$ by factor $\frac{L_s}{L_{s}-L_{u}} \geq 1$. Note that coded caching achieves the same scaling factor as uniform uncoded caching with $q_{b,l_s}^{\mathrm c} = L_u / L_s, \forall l_s \in \mathbb{L}_s$, where we have $\frac{1}{q_{b}^{\mathrm{a}}} = \frac{L_s}{L_{s}-L_{u}} \le \frac{1}{q_{b}^{\mathrm{a},*}}$, cf. \eqref{eq46_opt}.

In \cite{Maddah-Ali2014}, the delivery of requested but uncached (sub)files is divided into a number of multicast transmissions, each serving a group of $t+1$ users at a time. This strategy may undermine the spatial multiplexing gains provided by massive MIMO and limit the system performance, especially when the cache size is small. To tackle this issue, in \cite[Section V]{Ngo2018}, the authors employed massive MIMO for simultaneous delivery of a subfile to each user within a cell, such that the offloading gain of coded caching and the multiplexing gain of massive MIMO can be jointly exploited. Herein, different from \cite{Maddah-Ali2014,Ngo2018}, we combine coded caching with the proposed cache-aided massive MIMO to deliver one subfile to each of the $K$ users within a cell at a time. For convenience of presentation, the subfiles intended for a user are randomly permuted before being sequentially delivered. Compared with the scheme investigated in \cite[Section V]{Ngo2018}, which is henceforth referred to as Baseline Scheme~2, our proposed scheme exploits the pilot decontamination and interference cancellation enabled by coded caching for additional performance gains.

Since all users have cached the same number, i.e., $\frac{L_u}{L_s}C_K^t$, of subfiles of each file, the performance of coded caching is independent of the requesting probabilities. Moreover, thanks to the above described cache placement, users $(j,l), \forall j \in \mathbb{B}$, have the same cache status. Hence, we denote the probabilities that user $(j,l)$ causes interference to and is interfered by user $(b,k)$ by $p_{k,l}^{\mathrm{i}}$ and $p_{k,l}^{\mathrm{n}}$, respectively, where both probabilities are independent of the cell indices.

\vspace{-1em}
\begin{lem} \label{lem4}
\emph{
For the considered coded caching, $p_{k,l}^{\mathrm{i}}$ and $p_{k,l}^{\mathrm{n}}$ are given by
\begin{subequations}\label{eqlem4}
\begin{numcases}{p_{k,l}^{\mathrm{i}}=p_{k,l}^{\mathrm{n}}=}
1,\;\;\;\;\;\;\;\;\;\;\;\;\;\;\;\;\; &{\text{if }} $l = k$, \\
\tfrac{K-t-1}{K-1},\;\;\; &\text{otherwise}.
\end{numcases}
\end{subequations}
}
\end{lem}
\vspace{-0.5em}
\begin{IEEEproof}
Please refer to Appendix~\ref{lemma4}.
\end{IEEEproof}

\subsubsection{MRT Precoding} \label{MRT-2}
With coded caching, each file is only partially cached at all users. However, as the placement of each subfile is binary, Proposition~\ref{prop1} is still applicable for deriving the equivalent ergodic rate of coded caching, as shown in the following lemma.
\vspace{-1em}
\begin{lem} \label{lem6}
\emph{
For jointly coded caching and MRT, a lower bound of the ECDR achieved at user $(b,k)$ when $K,M\to\infty$ with constant ratio $\rho_0$, is given as follows
\begin{equation}\label{eq11.2}
\mathcal{\tilde{R}}_{b,k}^{\mathrm{MRT,P2}} \!=\!\tfrac{L_s}{L_{\!s}-L_{u}} \!\! \log_2\left(\!1+\frac{\rho_0\hat{\beta}_{b,k,b}^{\mathrm{b}}} {p_{k,l}^{\mathrm{i}} \hat{\beta}_{b,k,b}^{\mathrm{b}}
\!+\! \sum\nolimits_{j \neq b} (p_{k,k}^{\mathrm{i}}\rho_0+p_{k,l}^{\mathrm{i}})\hat{\beta}_{b,k,j}^{\mathrm{b}}
\!+\! \sum\nolimits_{j \in \mathbb{B}} p_{k,l}^{\mathrm{i}}\tilde{\beta}_{b,k,j}^{\mathrm{b}} \!+\! {1}/{E_0}}\right).
\end{equation}
}
\end{lem}
\vspace{-0.5em}
\begin{IEEEproof}
The result follows from \eqref{eq6.6} and \eqref{eq6.7}, along a logical line similar to the proof of Lemma~\ref{lem5}. However, for coded caching, we have $E_{j,l}=E_0/K$ as all users are active, and $F/L_d=L_s/(L_s-L_u)$. Moreover, exploiting the LLN, the number of users in set $\mathbb{U}_{b,k,j}$ converges to $p_{k,l}^{\mathrm{i}}K$ with probability one for $K \to \infty$. Using this fact, each term in the denominator of \eqref{eq6.7} can be calculated similar to \eqref{eq10.1} and the result in \eqref{eq11.2} is obtained.
\end{IEEEproof}

By setting $p_{k,l}^{\mathrm{i}}=1,j\in \mathbb{B}$, in \eqref{eq11.2}, a lower bound on the ECDR 
of user $(b,k)$ for Baseline Scheme~$2$ is given as follows
\begin{equation}\label{eq11.3}
\mathcal{\tilde{R}}_{b,k}^{\mathrm{MRT,B2}}=\tfrac{L_s}{L_{s}-L_{u}}\log_2\left(1+ \frac{\rho_0\hat{\beta}_{b,k,b}^{\mathrm{b}}} {\hat{\beta}_{b,k,b}^{\mathrm{b}}
+ \sum\nolimits_{j \neq b}  (\rho_0+1)\hat{\beta}_{b,k,j}^{\mathrm{b}}
+ \sum\nolimits_{j \in \mathbb{B}} \tilde{\beta}_{b,k,j}^{\mathrm{b}} + {1}/{E_0}} \right).
\end{equation}
Compared with Baseline Scheme~$2$, the proposed scheme exploits coded caching to reduce the interference for massive MIMO transmission. Therefore, we have $\mathcal{\tilde{R}}_{b,k}^{\mathrm{MRT,P2}} \geq \mathcal{\tilde{R}}_{b,k}^{\mathrm{MRT,B2}}$.

\subsubsection{ZF Precoding} \label{ZF-2}
A lower bound on the ECDR for joint coded caching and ZF precoding is given by
\begin{equation}\label{eq11.4}
\tilde{\mathcal{R}}_{b,k}^{\mathrm{ZF,P2}} \!=\! \tfrac{L_s}{L_{s}-L_{u}}\log_2 \!\left(\!1\!+\!\frac{\left(\rho_0-p_{k,l}^{\mathrm{n}}\right)\hat{\beta}_{b,k,b}^{\mathrm{b}}}
{\sum\nolimits_{j\neq b}\left((\rho_0-p_{k,l}^{\mathrm{n}})p_{k,k}^{\mathrm{i}} \!+\! (1-p_{k,l}^{\mathrm{i}})\right)\hat{\beta}_{b,k,j}^{\mathrm{b}}
\!+\! \sum\nolimits_{j \in \mathbb{B}} p_{k,l}^{\mathrm{i}}\tilde{\beta}_{b,k,j}^{\mathrm{b}} \!+\! {1}/{E_0}}\!\right),
\end{equation}
which follows from \eqref{eq7.8} using the same arguments as for Lemma~\ref{lem6}. A lower bound of the ECDR for Baseline Scheme~$2$ is obtained from \eqref{eq11.4} by setting $p_{k,l}^{\mathrm{n}} = p_{k,l}^{\mathrm{c}} =1$ and is given by 
\vspace{-0.3em}
\begin{equation}\label{eq11.5}
\tilde{\mathcal{R}}_{b,k}^{\mathrm{ZF,B2}}=\tfrac{L_s}{L_{s}-L_{u}}\log_2\left(1+\frac{(\rho_0-1)\hat{\beta}_{b,k,b}^{\mathrm{b}}}
{\sum\nolimits_{j\neq b}(\rho_0-1)\hat{\beta}_{b,k,j}^{\mathrm{b}}
+\sum\nolimits_{j \in \mathbb{B}} \tilde{\beta}_{b,k,j}^{\mathrm{b}}+{1}/{E_0}}\right).
\end{equation}
Compared with Baseline Scheme~$2$, the proposed scheme can: i) enhance the power of user $(b,k)$ since the interference from the other users can be partially mitigated as $p_{k,l}^{\mathrm{n}}\leq 1$, which is enabled by coded caching without having to rely on ZF precoding, and ii) reduce the intra-cell interference by a factor of $p_{k,l}^{\mathrm{i}}\leq 1$. Hence, we have $\tilde{\mathcal{R}}^{\mathrm{ZF,P2}} \geq \tilde{\mathcal{R}}^{\mathrm{ZF,B2}}$.

\subsubsection{RZF Precoding} \label{RZF-2}
By adopting identical regularization parameters for all users, the ECDR for joint coded caching and RZF precoding is given as follows
\begin{equation}\label{eq11.6}
\!\mathcal{R}_{b,k}^{\mathrm{RZF,P2}}\!=\!\tfrac{L_s}{\!L_{\!s}\!-\!L_{\!u}}
\!\log_2\!\left(\!1 + \!\frac{{\rho_{0} \hat{\beta}_{b,k,b}^{\mathrm{b}} \mathcal{G}^2_{b} } / { \overline{\mathcal{G}}_{b} }}{\!\! \frac{p_{k,l}^{\mathrm{i}}\hat{\beta}_{b,k,b}^{\mathrm{b}}}{\left(1+\mathcal{G}_{b} \right)^2}\!+\!\! \sum\limits_{j\neq b}\!\hat{\beta}_{b,k,j}^{\mathrm{b}}\!\!\left(\!\!\frac{\rho_{0} p_{k,k}^{\mathrm{i}} \mathcal{G}^2_{j} }{ \overline{\mathcal{G}}_{j} }
\!+\!\frac{p_{k,l}^{\mathrm{n}}p_{k,l}^{\mathrm{i}}}{\left(1\!+\!\mathcal{G}_{j} \right)^2}\!+\!(1\!-\!p_{k,l}^{\mathrm{n}})p_{k,l}^{\mathrm{i}}\!\!\right) \!+\!\sum\limits_{j \in \mathbb{B}} \! p_{k,l}^{\mathrm{i}}\tilde{\beta}_{b,k,j}^{\mathrm{b}}\!+\!\frac{1}{E_0}}\!\right)\!.
\end{equation}
Eq. \eqref{eq11.6} follows from \eqref{eq8.9} and the fact that the numbers of users in sets $\mathbb{U}_{b,k,j}$ and $\mathbb{N}_{k,j}$ in \eqref{eq8.9} converge to $p_{k,l}^{\mathrm{i}}K$ and $p_{k,l}^{\mathrm{n}}K$, respectively, when $M, K \rightarrow \infty$ but $\rho_0$ is fixed. Based on \eqref{eq11.6}, the ECDR for Baseline Scheme~$2$ is given as follows
\begin{equation}\label{eq11.7}
\mathcal{R}_{b,k}^{\mathrm{RZF,B2}}\!=\!\tfrac{L_s}{\!L_{\!s}\!-\!L_{\!u}}\log_2\!\left(\!1\!+\!\frac{{\rho_{0} \hat{\beta}_{b,k,b}^{\mathrm{b}} \mathcal{G}_{0}^2 } / { \overline{\mathcal{G}}_{0} }}{\frac{\hat{\beta}_{b,k,b}^{\mathrm{b}}}{\left(1+\mathcal{G}_{0} \right)^2}\!+\! \sum\nolimits_{j\neq b}\!\hat{\beta}_{b,k,j}^{\mathrm{b}}\!\left(\!\frac{\rho_{0}\mathcal{G}_{0}^2}{ \overline{\mathcal{G}}_{0} }
\!+\!\frac{1}{\left(1+\mathcal{G}_{0} \right)^2}\!\right) \!+\!\sum\nolimits_{j\in \mathbb{B}} \tilde{\beta}_{b,k,j}^{\mathrm{b}}\!+\!\frac{1}{E_0}}\!\right).
\end{equation}
We have $\mathcal{R}_{b,k}^{\mathrm{RZF,P2}} \geq \mathcal{R}_{b,k}^{\mathrm{RZF,B2}}$ since, by employing the proposed scheme, the effective intra-cell and inter-cell interference can be reduced by factors $p_{k,l}^{\mathrm{i}}<1$ and $p_{k,l}^{\mathrm{n}}<1$, respectively.

\vspace{-0.6em}
\section{Performance Evaluation} \label{Simulation Results}
\vspace{-0.1em}
In this section, we evaluate the performance of the proposed cache-aided massive MIMO scheme by simulations. Let $\mathrm{SNR} \triangleq 10\log_{10}E_0$ be the transmit SNR. The pathloss is modeled as $\beta_{b,k,j}=\frac{1}{1+d_{b,k,j}^\gamma}$, where $d_{b,k,j}$ is the distance between BS $j$ and user $(b,k)$ and $\gamma$ is the pathloss exponent  \cite{Adhikary2013}. Moreover, we normalize the cell radius such that $d_{b,k,j}<1$ for $b=j$ and $d_{b,k,j}=1+U$ for $b\neq j$, where $U\!\in[0,1]$ is a random variable \cite{Wen2015}. To maximize the performance of RZF precoding, the regularization parameter for all users, $\alpha$, is optimized numerically for each parameter setting. The other relevant system parameters are provided in Table \ref{table1}.
\vspace{-0.5em}
\begin{table}[h]
\vspace{-0.5em}
\renewcommand\arraystretch{1.2}
\setlength{\tabcolsep}{0.025\columnwidth}				
\normalsize
\centering
\caption{Simulation Parameters\vspace{-1em}}			
\begin{tabular}{|c|c|c|c|c|c|c|c|}			
	\hline
	Parameter & $B$ &$L_s$ & $F$ & $\tau$ & $p$ & $\gamma$ & ($\eta_1,\eta_2,\ldots,\eta_B$)  \\
	\hline
	Value & $3,4$ &$100$  & $1 \;{\rm MByte}$ & $K$ & $1$ & $3.8$ & ($0.6,0.5,0.4,0.3$)  \\
	\hline					
\end{tabular}		
\label{table1}
\end{table}
\vspace{-1em}

\begin{figure}[!tbp]
\vspace{-1em}
\centering
\begin{minipage}[b]{0.485\linewidth}
\centering
\includegraphics[scale=0.56]{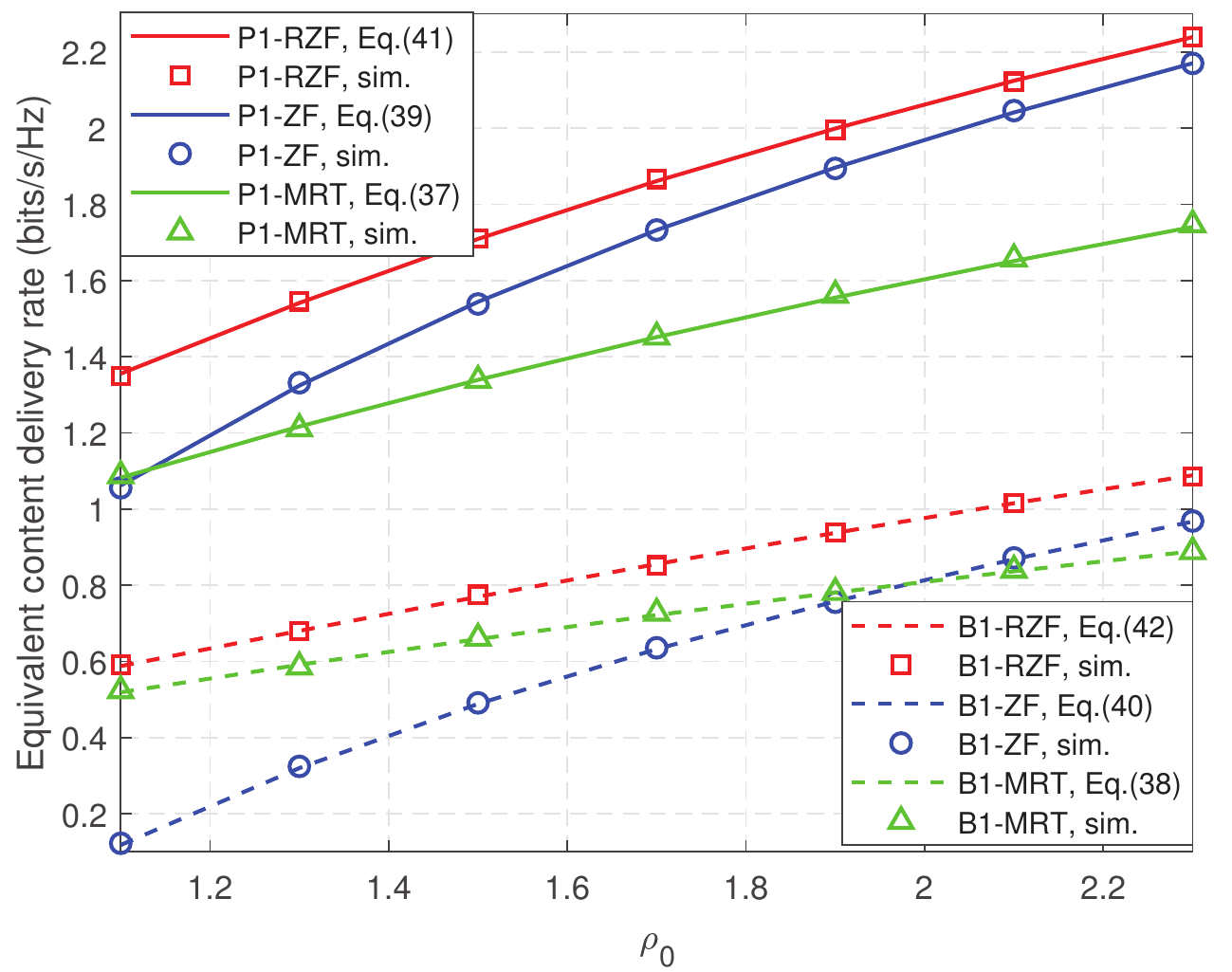}
\vspace{-1em}
\caption{Equivalent content delivery rate for MRT, ZF, and RZF precoding versus number of BS antennas per user, $\rho_0$. Uncoded caching, $L_u=6$, $B=3$, $K=200$, and $\mathrm{SNR}=20\;{\mathrm{dB}}$.}
\vspace{-2em}
\label{fig1}
\end{minipage}
\hspace{0.2em}
\begin{minipage}[b]{0.485\linewidth}
\centering
\includegraphics[scale=0.55]{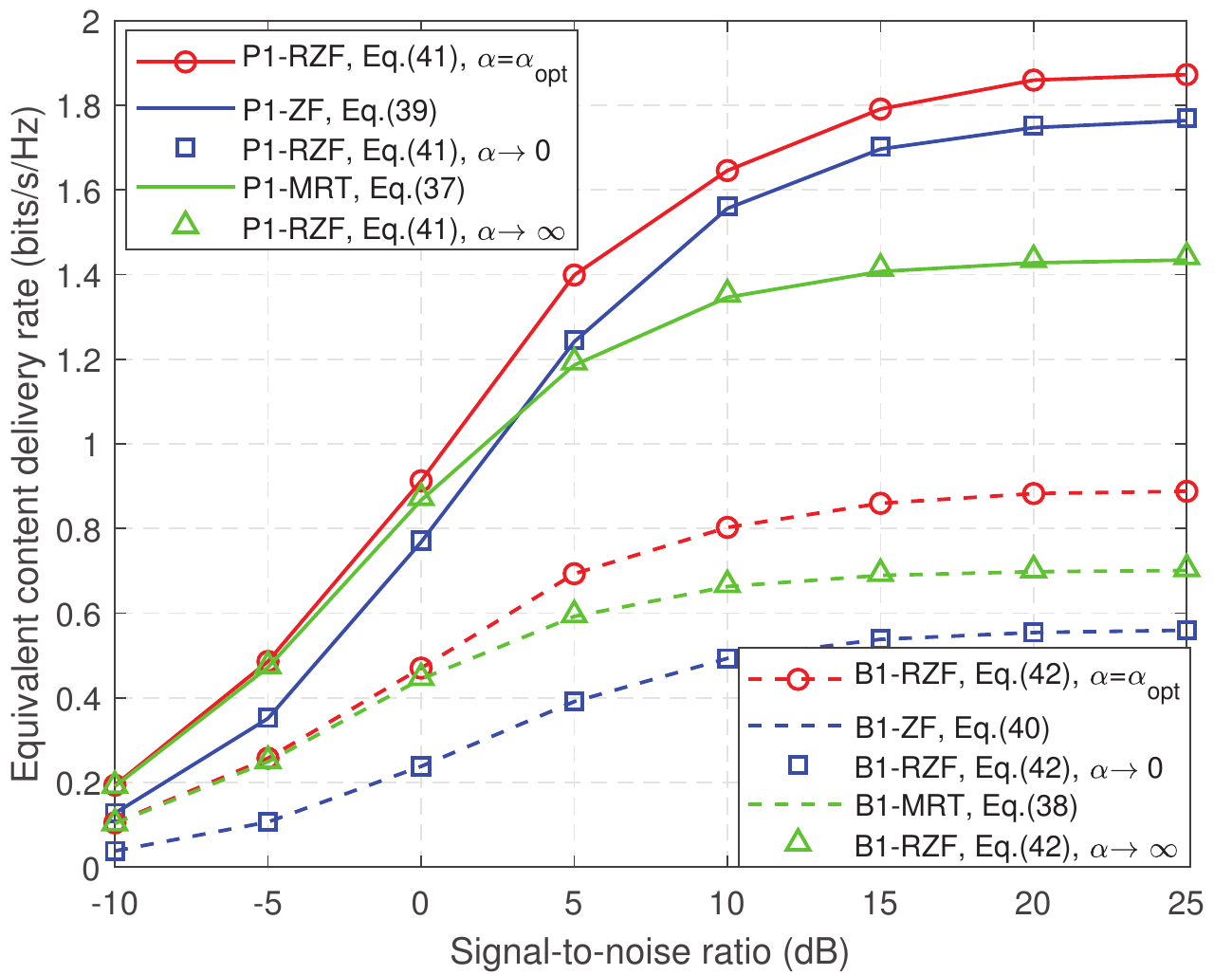}
\vspace{-1em}
\caption{Equivalent content delivery rate for MRT, ZF, and RZF precoding versus transmit SNR. Uncoded caching, $B=3$, $K=200$, $\rho_0=1.5$, and $L_u=6$.}
\label{fig2}
\end{minipage}
\vspace{-2em}
\end{figure}

We first evaluate the performance of cache-aided massive MIMO under uncoded caching, where each user $(b,k)$ caches file $l_s$ with probability $q_{b,l_s}^{\mathrm{c}}$ in \eqref{eq46_opt}. Fig.~\ref{fig1} shows the ECDRs for MRT, ZF, and RZF precoding as functions of the number of BS antennas per user, $\rho_0$, where `$\rm P1$' and `$\rm B1$' refer to the proposed scheme and Baseline Scheme~$1$, respectively. Both analytical and Monte Carlo simulation results are presented in Fig.~\ref{fig1}. The analytical expressions of the ECDR for the proposed scheme and Baseline Scheme~$1$ are given by \eqref{eq10.1}$-$\eqref{eq10.6} and the simulation results are averaged over $10^5$ realizations of the distances between users and BSs, channel fading, and user requests. As can be observed, the analytical and simulation results are in good agreement, which validates the derivations in Sections III and IV. Moreover, for all considered precoders, the proposed scheme achieves significantly higher ECDRs than Baseline Scheme~$1$, particularly when $\rho_0$ becomes large. This is because, on the one hand, caching offloads the cellular traffic of inactive users and mitigates the multiuser interference for active users. On the other hand, with the proposed scheme, caching is further exploited for BS precoding to increase the received signal power and/or the spatial degrees of freedom, which further improves the ECDR. For example, for MRT precoding, the proposed scheme at $\rho_0 =  1.1$ achieves even higher performance than Baseline Scheme~$1$ at $\rho_0 = 2.2$. From Fig.~\ref{fig1} we also observe that the proposed scheme achieves the largest performance gains over Baseline Scheme~$1$ for ZF precoding, even when the number of antennas approaches the number of users. This is because, for a small $\rho_0$, the signal space for ZF precoder design is severely constrained. The proposed scheme exploits cache-enabled interference cancellation to reduce the number of ZF precoding constraints. Consequently, ZF precoding can benefit from the additional spatial degrees of freedom enabled by caching, leading to a significant performance gain. Moreover, by balancing interference cancellation and power enhancement, the proposed scheme with RZF precoding achieves the best performance. These results confirm that the joint design of caching at the users and linear precoding at the BSs can substantially enhance the performance of massive MIMO systems for different numbers of BS antennas per user, $\rho_0$.

Fig.~\ref{fig2} illustrates the ECDRs for MRT, ZF, and RZF precoding as functions of the transmit SNR. From Fig.~\ref{fig2} we observe that, as expected, RZF precoding achieves the same performance as MRT and ZF precoding when $\alpha\rightarrow 0$ and $\alpha\rightarrow \infty$, respectively. Moreover, for all considered precoders, the proposed scheme significantly outperforms Baseline Scheme~$1$, especially in the high SNR regime. For example, for MRT precoding, the proposed scheme achieves an SNR gain of more than $9\;\mathrm{dB}$ at an ECDR of $0.6\;\text{bit/s/Hz}$ compared with Baseline Scheme~$1$. This performance improvement occurs because the system is severely interference-limited in the high SNR regime, and with the proposed scheme, the interference can be mitigated by exploiting caching. Additionally, for ZF precoding, the performance gain enabled by offloading and interference cancellation is also significant in the low SNR regime. For example, for ZF precoding, the proposed scheme achieves an SNR gain of more than $6\;\mathrm{dB}$ over Baseline Scheme~$1$ at an ECDR of $0.2\;\text{bit/s/Hz}$.

\begin{figure}[!tbp]
\vspace{-1em}
\centering
\begin{minipage}[t]{0.485\linewidth}
\centering
\includegraphics[scale=0.54]{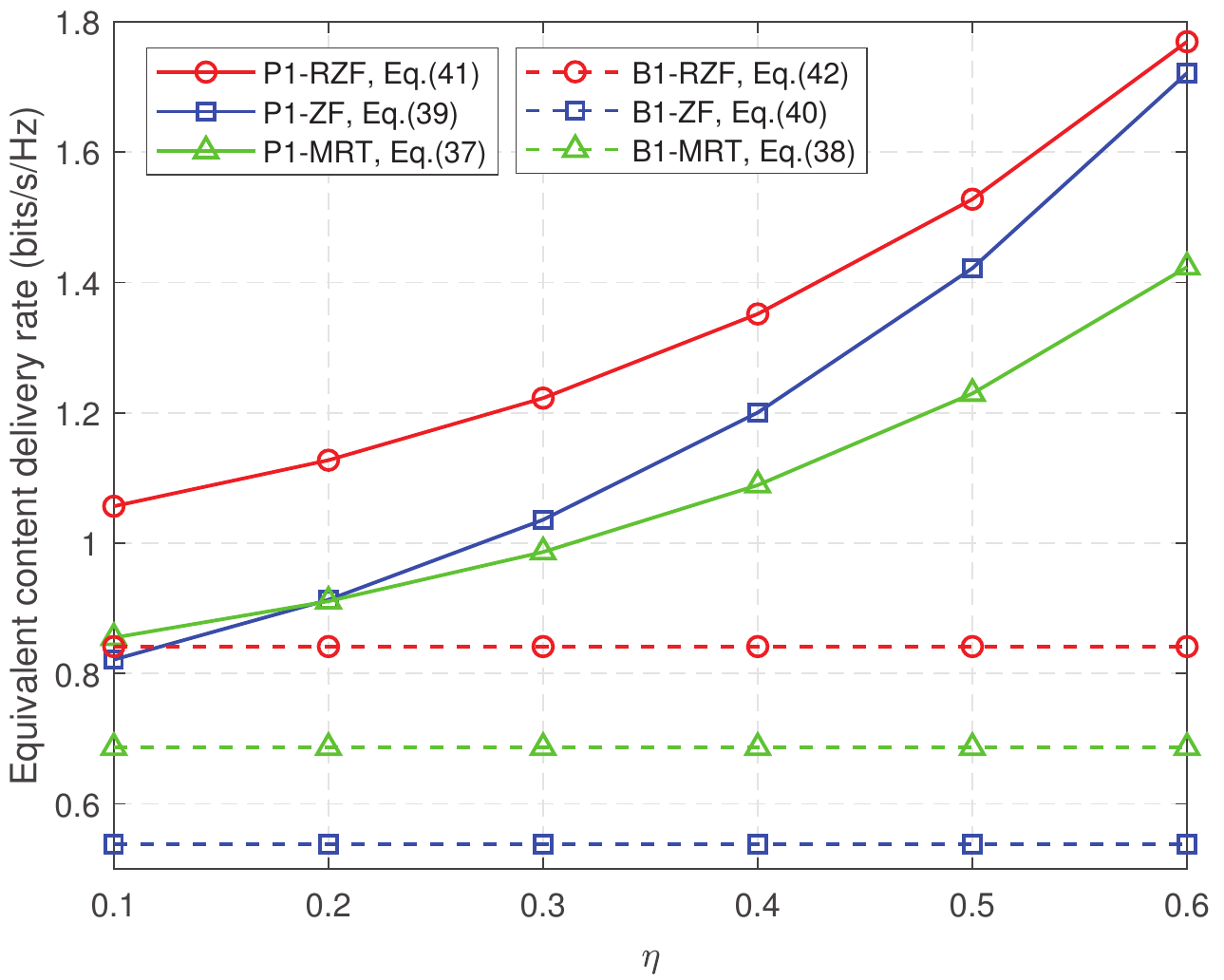}
\vspace{-1em}
\caption{Equivalent content delivery rate for MRT, ZF, and RZF precoding versus Zipf exponent, $\eta_b=\eta$. Uncoded caching, $B=3$, $K=200$, $\rho_0=1.5$, $L_u=6$, and $\mathrm{SNR}=20\;{\mathrm{dB}}$.}
\label{fig4}
\end{minipage}
\hspace{0.2em}
\begin{minipage}[t]{0.485\linewidth}
\centering
\includegraphics[scale=0.55]{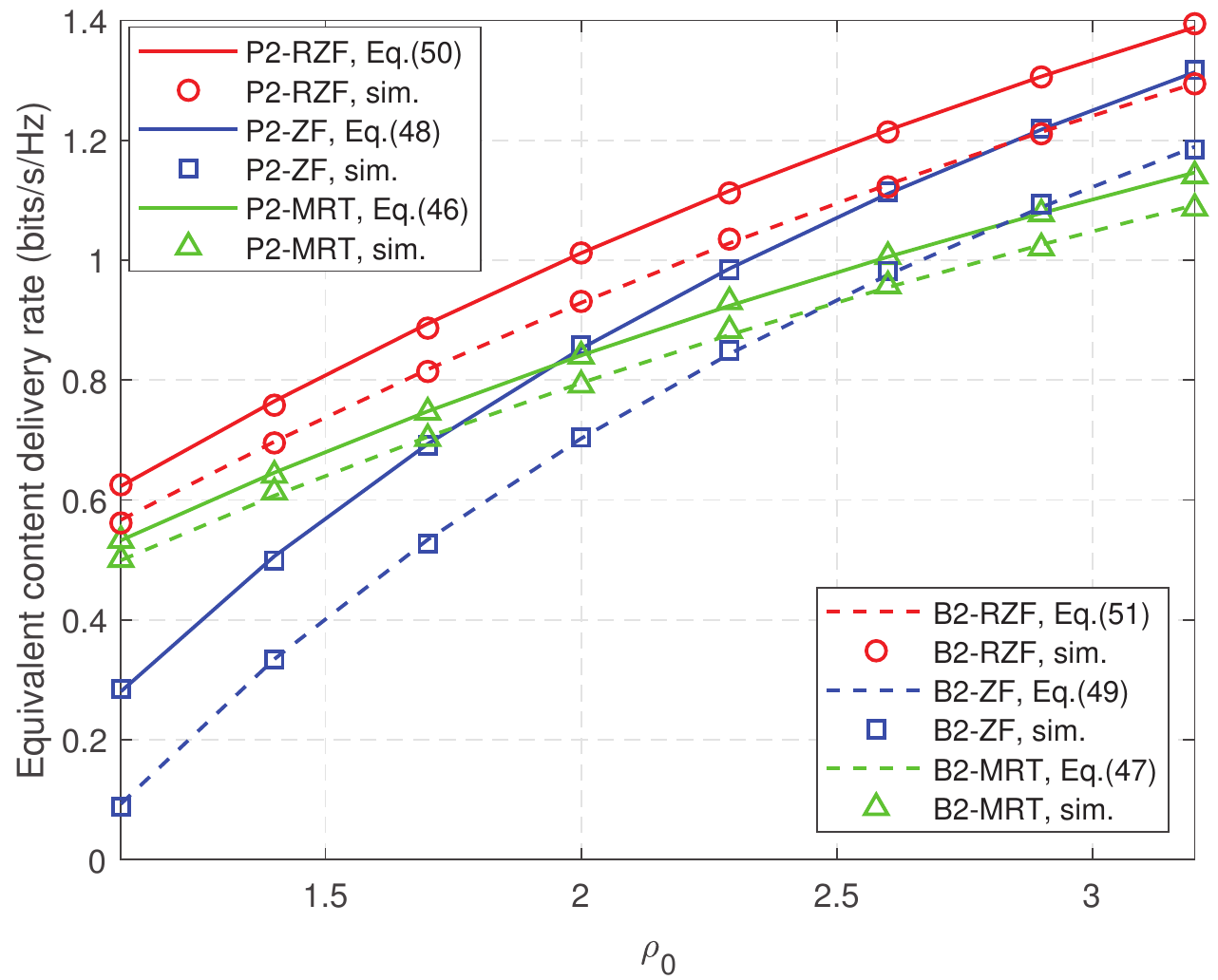}
\vspace{-1em}
\caption{Equivalent content delivery rate for MRT, ZF, and RZF precoding versus number of BS antennas per user, $\rho_0$. Coded caching, $L_u=20$, $B=4$, $K=100$, and $\mathrm{SNR}=10\;{\mathrm{dB}}$.\vspace{-2em}}
\label{fig6}
\end{minipage}
\vspace{-1em}
\end{figure}

Fig.~\ref{fig4} shows the ECDR for MRT, ZF, and RZF precoding as a function of the Zipf exponent, $\eta_b=\eta$. We observe that, for all considered precoders, the ECDR of the proposed scheme increases monotonically with $\eta$. This is because the users' requests become highly concentrated around the most popular files as $\eta$ increases, cf. \eqref{eq9.1}, such that caching enables more offloading and interference cancellation opportunities for the proposed scheme.
In contrast, these caching gains are not available for Baseline Scheme~$1$, whereby the ECDR of Baseline Scheme~$1$ is independent of $\eta$.

Next, we consider coded caching and compare the performance of the proposed cache-aided MIMO scheme with Baseline Scheme~$2$. The performance of the proposed scheme and Baseline Scheme~$2$ is evaluated analytically based on \eqref{eq11.2}$-$\eqref{eq11.7}. Fig.~\ref{fig6} shows the ECDR for MRT, ZF, and RZF precoding as functions of the number of BS antennas per user, $\rho_0$, where `$\rm P2$' and `$\rm B2$' denote the proposed scheme with coded caching and Baseline Scheme~$2$, respectively. From Fig.~\ref{fig6} we observe that, for all considered precoders, the proposed scheme achieves significantly higher delivery rates than Baseline Scheme~$2$. This performance enhancement occurs because, compared with Baseline Scheme~$2$, the proposed scheme exploits joint caching and precoding to increase the available spatial degrees of freedom via interference cancellation and to increase the received signal power, which enhances the ECDR also for coded caching.

\vspace{-0.5em}
\section{Conclusions} \label{Conclusion} 
In this paper, a novel cache-aided massive MIMO scheme was proposed for downlink data transmission in multi-cell cellular networks. In addition to reaping the advantages of caching and massive MIMO, the proposed scheme also facilitates interference cancellation at the receivers. Exploiting both cache-enabled offloading and interference cancellation, enhanced uplink channel estimation and downlink MRT, ZF, and RZF precoders were proposed to increase the transmit power per user and to reduce the impairments caused by intra- and inter-cell interference and imperfect CSI.
For a given arbitrary cache placement, closed-form expressions for the ECDRs 
of cache-aided massive MIMO were presented for MRT, ZF, and RZF precoding. Subsequently, the derived results were specialized to (random) uncoded caching and coded caching. Both analytical and simulation results show that cache-aided massive MIMO significantly improves the performance of linear precoding techniques for massive MIMO, even when the number of BS antennas per user is small.

\vspace{-0.8em}
\begin{appendix}
\subsection{Proof of Proposition~\ref{prop1}} \label{proof1}
Substituting \eqref{eq6.3} into \eqref{eq5.3}, we have $\mathcal{R}_{b,k}^{\mathrm{MRT}}\geq \frac{F}{L_{d}}\log_2\left(1+\left(\mathcal{E}\left\{(\gamma_{b,k}^{\mathrm{MRT}})^{-1}\right\}\right)^{-1}\right)$, where
\vspace{-0.5em}
\begin{equation}\label{eqp1.1}
\vspace{-0.5em}
\mathcal{E}\!\left\{\tfrac{1}{\gamma_{b,k}^{\mathrm{MRT}}}\right\}
\!=\!\!\!\!\!\sum\limits_{(j,l) \in \mathbb{U}_{b,k}} \!\!
\tfrac{\!E_{j,l}\lambda_{j,l}^{\mathrm{MRT}}\!}{\! E_{b,k}\lambda_{b,k}^{\mathrm{MRT}}}
\mathcal{E}\!\left\{\!\! \tfrac{|  {\hat{\bf{h}}}_{b,k,j}^{\mathrm H}{\hat{\bf{h}}}_{j,l,j} |^2}{| {\hat{\bf{h}}}_{b,k,b}^{\mathrm H}{\hat{\bf{h}}}_{b,k,b} |^2} \!+\!
\tfrac{| {\tilde{\bf{h}}}_{b,k,j}^{\mathrm H}{\hat{\bf{h}}}_{j,l,j}|^2}{| {\hat{\bf{h}}}_{b,k,b}^{\mathrm H}{\hat{\bf{h}}}_{b,k,b} |^2} \!\right\} \!
+ \mathcal{E}\!\left\{\! \tfrac{| {\tilde{\bf{h}}}_{b,k,b}^{\mathrm H}{\hat{\bf{h}}}_{b,k,b}|^2}{| {\hat{\bf{h}}}_{b,k,b}^{\mathrm H}{\hat{\bf{h}}}_{b,k,b} |^2} \!\right\} \!
+\!\tfrac
{\mathcal{E}\left\{\! {1}/{| {\hat{\bf{h}}}_{b,k,b}^{\mathrm H}{\hat{\bf{h}}}_{b,k,b} |^2} \!\right\}
}{E_{b,k} \lambda_{b,k}^{\mathrm{MRT}}}.\!\! 
\end{equation}
As ${\hat{\bf{h}}}_{b,k,j}^{\mathrm H}{\hat{\bf{h}}}_{j,l,j}$ and $ {\hat{\bf{h}}}_{b,k,b}^{\mathrm H}{\hat{\bf{h}}}_{b,k,b} $ are uncorrelated for $j\neq b$ and $\mathcal{E} \{ \tfrac{1}{4} {\hat{\beta}_{b,k,b}^2}/{| {\hat{\bf{h}}}_{b,k,b}^{\mathrm H}{\hat{\bf{h}}}_{b,k,b} |^2} \} $ corresponds to the second-order moment of an inverse chi-square distribution with $2M$ degrees of freedom \cite[Appendix~A]{Wei2019}, we have 
\vspace{-0.5em}
\begin{equation}\label{eqp1.2} 
\mathcal{E}\!\left\{\! {|  {\hat{\bf{h}}}_{b,k,j}^{\mathrm H}{\hat{\bf{h}}}_{j,l,j} |^2} / {| {\hat{\bf{h}}}_{b,k,b}^{\mathrm H}{\hat{\bf{h}}}_{b,k,b} |^2} \!\right\} =  
 \left( {(M\!-\!1)(M\!-\!2)\hat{\beta}_{b,k,b}^2} \right)^{-1} \mathcal{E}\!\left\{\! {|  {\hat{\bf{h}}}_{b,k,j}^{\mathrm H}{\hat{\bf{h}}}_{j,l,j} |^2} \!\right\},\;\;\;\; j\neq b. 
\end{equation} 
If $j\neq b$ and $l \neq k$, i.e., ${\hat{\bf{h}}}_{b,k,j}$ and ${\hat{\bf{h}}}_{j,l,j}$ are independent, we have $\mathcal{E}\{ {|  {\hat{\bf{h}}}_{b,k,j}^{\mathrm H}{\hat{\bf{h}}}_{j,l,j} |^2} \}
=M\hat{\beta}_{b,k,j}\hat{\beta}_{j,l,j}$ due to \cite[Eq. (5)]{N2013}.
Otherwise, i.e., if $j\neq b$ and $l = k$, we have ${\hat{\bf{h}}}_{b,k,j} = \tfrac{\beta_{b,k,j}}{\beta_{j,k,j}} {\hat{\bf{h}}}_{j,k,j}$, whereby $\mathcal{E}\{ {| {\hat{\bf{h}}}_{b,k,j}^{\mathrm H}{\hat{\bf{h}}}_{j,l,j} |^2} \}$ $=M(M+1)\hat{\beta}_{b,k,j}\hat{\beta}_{j,k,j}$ corresponds to the second-order moment of a chi-square distribution with $2M$ degrees of freedom \cite[Appendix~A]{Wei2019}. 
Consequently, this leads to
\vspace{-0.5em}
\begin{subequations}\label{eqp1.2a}
\begin{numcases}
{\mathcal{E}\!\left\{ \frac{|  {\hat{\bf{h}}}_{b,k,j}^{\mathrm H}{\hat{\bf{h}}}_{j,l,j} |^2}{| {\hat{\bf{h}}}_{b,k,b}^{\mathrm H}{\hat{\bf{h}}}_{b,k,b} |^2} \right\}
 = }
 {\hat{\beta}_{b,l,b}}{\left( (M-1)\hat{\beta}_{b,k,b} \right)^{-1}},
 &{\text{if }} $j = b$, \label{eq51a} \\
 {M\hat{\beta}_{b,k,j}\hat{\beta}_{j,l,j}\!}{\left( (M\!-\!1)(M\!-\!2)\hat{\beta}_{b,k,b}^2 \right)^{-1}}, 
&{\text{if }} $j \neq b,\; l\neq k$, \\
 {M(M+1)\hat{\beta}_{b,k,j}\hat{\beta}_{j,k,j} \!}{\left( (M\!-\!1)(M\!-\!2)\hat{\beta}_{b,k,b}^2 \right)^{-1}}, 
&\text{otherwise},    
\end{numcases}
\end{subequations}
where \eqref{eq51a} is due to \cite[Appendix A]{N2013}. Following a similar approach as above, we can further show
\vspace{-0.5em}
\begin{equation}\label{eqp1.8}
\vspace{-0.5em}
\mathcal{E}\left\{ \frac{| {\tilde{\bf{h}}}_{b,k,j}^{\mathrm H}{\hat{\bf{h}}}_{j,l,j}|^2}{| {\hat{\bf{h}}}_{b,k,b}^{\mathrm H}{\hat{\bf{h}}}_{b,k,b} |^2} \right\}
=\begin{cases}
{M \tilde{\beta}_{b,k,j}\hat{\beta}_{j,k,j}}{\left( (M\!-\!1)(M\!-\!2) \hat{\beta}_{b,k,b}^2 \right)^{-1}}, &{\text{if }} (j,l) \neq (b,k), \\
{\tilde{\beta}_{b,k,b}} {\left((M\!-\!1)\hat{\beta}_{b,k,b} \right)^{-1}}, &\text{otherwise}.
\end{cases}
\end{equation}
Substituting \eqref{eqp1.2}--\eqref{eqp1.8} and $\lambda_{j,l}^{\mathrm{MRT}}= ({M\hat{\beta}_{j,l,j}})^{-1}$ into \eqref{eqp1.1}, Proposition~\ref{prop1} is thus proved.

\vspace{-1em}
\subsection{Proof of Lemma~\ref{lemma1}} \label{lemmaproof1}
To satisfy the transmit power constraint ${\mathcal E} \{\| {\bf{w}}_{b,k}^{\mathrm{ZF}}\|^2 \}=1$, we require
\vspace{-0.5em}
\begin{equation}\label{eqp2.1}
\vspace{-0.5em}
\lambda_{b,k}^{\mathrm{ZF}}={1}/{\mathcal{E}\left\{ \| {{\bf{Q}}_{b,k}({\bf{Q}}_{b,k}^{\rm H}{\bf{Q}}_{b,k})^{-1}{\bf{e}}_1} \|^2 \right\}}.
\end{equation}
Note that $({{\bf{Q}}_{b,k}^{\rm H}{\bf{Q}}_{b,k}})^{-1}$ is a complex inverse Wishart matrix with $N_{b,k}^{\mathrm{n}}+1$ degrees of freedom and mean given by \cite[Appendix~B]{Wei2019} 
\vspace{-0.5em}
\begin{equation}
\mathcal{E}\left\{ ({{\bf{Q}}_{b,k}^{\rm H}{\bf{Q}}_{b,k}})^{-1} \right\} ={{\mathrm{diag} \left([\hat{\beta}_{b,k,b}^{-1}, \hat{\beta}_{b,\mathbb{N}_{k} (1),b}^{-1}, \ldots, \hat{\beta}_{b,\mathbb{N}_{k} (N_{b,k}^{\mathrm{n}}),b}^{-1}] \right) }} / {\left( M-(N_{b,k}^{\mathrm{n}}+1) \right)}. \label{eqp2.2} 
\end{equation}
Moreover, as ${ \| {\bf{Q}}_{b,k}({\bf{Q}}_{b,k}^{\rm H}{\bf{Q}}_{b,k})^{-1}{\bf{e}}_1 \|^2}  =  {\bf{e}}_1^{\rm H}({\bf{Q}}_{b,k}^{\rm H}{\bf{Q}}_{b,k})^{-1}{\bf{e}}_1 = \left[ ({\bf{Q}}_{b,k}^{\rm H}{\bf{Q}}_{b,k})^{-1} \right]_{1,1}$, we have
\vspace{-0.5em}
\begin{equation}
\mathcal{E}\left\{ { \left\| {\bf{Q}}_{b,k}({\bf{Q}}_{b,k}^{\rm H}{\bf{Q}}_{b,k})^{-1}{\bf{e}}_1 \right\| ^2} \right\}
={1}/{\left( (M-N_{b,k}^{\mathrm{n}}-1)\hat{\beta}_{b,k,b} \right)}.  \label{eqp2.3}
\end{equation}
Then, substituting \eqref{eqp2.3} into \eqref{eqp2.1}, Lemma~\ref{lemma1} is proved.

\vspace{-1em}
\subsection{Proof of Proposition~\ref{prop2}} \label{proof2}
Based on \eqref{eq5.1}, the SINR of user $(b,k)$ for ZF precoding is
\vspace{-0.5em}
\begin{equation}\label{eqp3.1}
\vspace{-0.5em}
\gamma_{b,k}^{\mathrm{ZF}}=\frac{P_{b,k}^{\mathrm{s,ZF}}}{\sum\nolimits_{(j,l) \in \mathbb{U}_{b,k}\setminus \mathbb{U}_{b,k,b}\!}\! P_{b,k,j,l}^{\mathrm{i,ZF}} \!+\! \sum\nolimits_{(j,l) \in \mathbb{V}_{b,k}\!}\! P_{b,k,j,l}^{\mathrm{e,ZF}} \!+\! 1},
\end{equation}
where $P_{b,k}^{\mathrm{s,ZF}}=E_{b,k} \lambda_{b,k}^{\mathrm{ZF}}| {\hat{\bf{h}}}_{b,k,b}^{\mathrm H}{\bf{Q}}_{b,k}({\bf{Q}}_{b,k}^{\rm H}{\bf{Q}}_{b,k})^{-1}{\bf{e}}_1|^2$. $P_{b,k,j,l}^{\mathrm{i,ZF}}$ and $P_{b,k,j,l}^{\mathrm{e,ZF}}$ are defined in \eqref{eq7.6}.

As ${\hat{\mathbf{h}}}_{b,k,b} = {\mathbf{q}}_{b,k,1} = \mathbf{Q}_{b,k} \mathbf{e}_1$, we have ${ |{\hat{\mathbf{h}}}_{{b,k},b}^{\mathrm{H}}{\mathbf{Q}}_{b,k}\left({\mathbf{Q}}_{b,k}^{\mathrm{H}}{\mathbf{Q}}_{b,k}\right)^{-1}{\bf{e}}_1 |^2}= 1$ 
and
\vspace{-0.5em}
\begin{equation}\label{eqp3.3}
\vspace{-0.5em}
P_{b,k}^{\mathrm{s,ZF}}=E_{b,k} \lambda_{b,k}^{\mathrm{ZF}}.
\end{equation}
On the other hand, as ${\hat{\bf{h}}}_{b,k,b}$ is orthogonal to ${\bf{w}}_{b,l}^{\mathrm{ZF}}$ in \eqref{eq7.1}, $P_{b,k,b,l}^{\mathrm{i,ZF}}=0,l\neq k$.
Substituting $P_{b,k,b,l}^{\mathrm{i,ZF}}$ and $P_{b,k}^{\mathrm{s,ZF}}$ into \eqref{eqp3.1}, \eqref{eq7.5} and \eqref{eq7.6} in Proposition~\ref{prop2} are thus proved.

Now, substituting \eqref{eq7.5} into \eqref{eq5.3}, we have $\mathcal{R}_{b,k}^{\mathrm{ZF}}\geq \frac{F}{L_{d}}\log_2\left(1+\left(\mathcal{E}\left\{(\gamma_{b,k}^{\mathrm{ZF}})^{-1}\right\}\right)^{-1}\right)$,
where
\vspace{-0.5em}
\begin{equation}\label{eqp3.5}
\vspace{-0.5em}
{E_{b,k} \lambda_{b,k}^{\mathrm{ZF}}}\mathcal{E}\left\{(\gamma_{b,k}^{\mathrm{ZF}})^{-1}\right\}
=\sum\nolimits_{(j,l) \in \mathbb{U}_{b,k}\setminus \mathbb{U}_{b,k,b}\!}\! \mathcal{E} \{ P_{b,k,j,l}^{\mathrm{i,ZF}} \} \!+\! \sum\nolimits_{(j,l) \in \mathbb{V}_{b,k}\!}\! \mathcal{E}\{P_{b,k,j,l}^{\mathrm{e,ZF}} \} \!+\! 1.
\end{equation}
Here, we evaluate $P_{b,k,j,l}^{\mathrm{i,ZF}}$ for three cases: (i) If $l=k$, i.e., ${\hat{\bf{h}}}_{b,k,j}$ and ${\hat{\bf{h}}}_{j,l,j}$ are collinear, 
this leads to
\vspace{-0.5em}
\begin{equation} \label{eqp3.6}
\vspace{-0.5em}
\left|  {\hat{\bf{h}}}_{b,k,j}^{\mathrm H}{\bf{Q}}_{j,l}({\bf{Q}}_{j,l}^{\rm H}{\bf{Q}}_{j,l})^{-1}{\bf{e}}_1 \right|^2
=\tfrac{\beta_{b,k,j}^2}{\beta_{j,k,j}^2}\left|  {\hat{\bf{h}}}_{j,k,j}^{\mathrm H}{\bf{Q}}_{j,k}({\bf{Q}}_{j,k}^{\rm H}{\bf{Q}}_{j,k})^{-1}{\bf{e}}_1 \right|^2 
\stackrel{(a)}{=} \tfrac{\hat{\beta}_{b,k,j}}{\hat{\beta}_{j,k,j}},
\end{equation}
where $(a)$ follows similarly from \eqref{eqp3.3} and $\frac{\hat{\beta}_{j,k,j}}{\hat{\beta}_{b,k,j}}=\frac{{\beta}_{j,k,j}^2}{{\beta}_{b,k,j}^2}$, cf. Section \ref{Uplink Channel Estimation}.
(ii) If $l\neq k$ and ${\hat{\bf{h}}}_{j,k,j}\in {\bf{Q}}_{j,l}$, i.e., ${\hat{\bf{h}}}_{j,k,j}$ is orthogonal to ${\bf{w}}_{j,l}^{\mathrm{ZF}}$, we have ${\hat{\bf{h}}}_{b,k,j} = \frac{\beta_{b,k,j}}{\beta_{j,k,j}} {\hat{\bf{h}}}_{j,k,j}$, such that ${\hat{\bf{h}}}_{b,k,j}^{\mathrm H}{\bf{Q}}_{j,l}({\bf{Q}}_{j,l}^{\rm H}{\bf{Q}}_{j,l})^{-1}{\bf{e}}_1=0$, i.e., ${\hat{\bf{h}}}_{b,k,j}$ is orthogonal to ${\bf{w}}_{j,l}^{\mathrm{ZF}}$.
(iii) If $l\neq k$ and ${\hat{\bf{h}}}_{j,k,j}\notin {\bf{Q}}_{j,l}$, we have that ${\hat{\bf{h}}}_{b,k,j}$ and ${\bf{Q}}_{j,l}$ are independent, and
\vspace{-0.5em}
\begin{align}\label{eqp3.7} 
\mathcal{E}\left\{ {| {\hat{\bf{h}}}_{b,k,j}^{\mathrm H}{\bf{Q}}_{j,l}({\bf{Q}}_{j,l}^{\rm H}{\bf{Q}}_{j,l})^{-1}{\bf{e}}_1 |^2} \right\} 
&\!=\!\mathcal{E}_{{\bf{Q}}_{j,l}} \left\{\! \mathcal{E}_{{\hat{\bf{h}}}_{b,k,j}} 
\left\{\! {{\bf{e}}_1^{\mathrm{H}}({\bf{Q}}_{j,l}^{\rm H}{\bf{Q}}_{j,l})^{\!-\!1}{\bf{Q}}_{j,l}^{\mathrm H}{\hat{\bf{h}}}_{b,k,j} {\hat{\bf{h}}}_{b,k,j}^{\mathrm H}{\bf{Q}}_{j,l}({\bf{Q}}_{j,l}^{\rm H}{\bf{Q}}_{j,l})^{\!-\!1}{\bf{e}}_1 } \!\right\} \!\right\} \nonumber \\
&\!\stackrel{(b)}{=}\!{\hat{\beta}_{b,k,j}}/{\left( (M\!\!-\!\!N_{j,l}^{\mathrm{n}}\!\!-\!\!1)\!\hat{\beta}_{j,l,j} \right)}, 
\end{align}
where $(b)$ is due to \eqref{eqp2.2}.
Following a similar approach as in \eqref{eqp3.3}, we can also show that
\vspace{-0.5em}
\begin{equation} \label{eqp3.8}
\vspace{-0.5em}
P_{b,k,j,l}^{\mathrm{e,ZF}}={\tilde{\beta}_{b,k,j}}E_{j,l},
\end{equation}
as ${\tilde{\bf{h}}}_{b,k,j}$ and ${\bf{Q}}_{j,l}$ are independent. Finally, substituting \eqref{eqp3.6}$-$\eqref{eqp3.8} into \eqref{eqp3.5}, \eqref{eq7.8} and \eqref{eq7.9} in Proposition~\ref{prop2} are proved.

\vspace{-1em}
\subsection{Proof of Lemma~\ref{lemma2}} \label{lemmaproof2}
To satisfy $\mathcal{E} \left\{\| {\bf{w}}_{b,k}^{\mathrm{RZF}}\|^2 \right\}=1$, we require
\vspace{-0.5em}
\begin{equation}\label{eqp5.1}
\vspace{-0.5em}
1/\lambda_{b,k}^{\mathrm{RZF}}=\mathcal{E}\left\{ \| ({\bf{F}}_{b,k}^{\rm H}{\bf{F}}_{b,k}+\alpha_{b,k}{\bf{I}}_M)^{-1}{\bf{f}}_{b,k,1} \|^2 \right\}
=\mathcal{E}\left\{ \| ({\bf{F}}_{b,k}^{\rm H}{\bf{F}}_{b,k}+\alpha_{b,k}{\bf{I}}_M)^{-1}{\hat{\bf{g}}}_{b,k,b} \|^2 \right\}.
\end{equation}
Denote by ${\bf{F}}_{b,k(b,k)}$ the residual matrix obtained by deleting vector ${\hat{\bf{g}}}_{b,k,b}$ in ${\bf{F}}_{b,k}$. Moreover, define $X_{b,k}={\hat{\bf{g}}}_{b,k,b}^{\rm H} ({\bf{F}}_{b,k(b,k)}^{\rm H}{\bf{F}}_{b,k(b,k)}+\alpha_{b,k}{\bf{I}}_M)^{-1}{\hat{\bf{g}}}_{b,k,b}$ and ${\bf{\Phi}}_{b,k}=({\bf{F}}_{b,k(b,k)}^{\rm H}{\bf{F}}_{b,k(b,k)}+\alpha_{b,k}{\bf{I}}_M)^{-1}$. 
By applying the matrix inversion lemma \cite{Nguyen2008}, we have
\vspace{-0.5em}
\begin{equation}\label{eqp5.2}
\vspace{-0.5em}
({\bf{F}}_{b,k}^{\rm H}{\bf{F}}_{b,k}+\alpha_{b,k}{\bf{I}}_M)^{-1}{\hat{\bf{g}}}_{b,k,b}
={\left(1+X_{b,k}\right)^{-1} } {{\bf{\Phi}}_{b,k}{\hat{\bf{g}}}_{b,k,b}}.
\end{equation}
Then, in the large system limit, where $M, K \rightarrow \infty$ but $\rho_0 = M/K$ is finite and fixed, we have $\rho_{b,k}^{-1}= N_{b,k}^{\mathrm{n}} / M \le \rho_0^{-1}$ such that $\rho_{b,k}^{-1}$ is finite; consequently, $X_{b,k}$ converges (almost surely) to \cite{Nguyen2008, Evans2000} 
\vspace{-0.5em}
\begin{equation}\label{eqp5.5}
\vspace{-0.5em}
\lim_{M, K \rightarrow \infty} \frac{1}{M} \mathrm{tr} \left( \mathbf{\Phi}_{b,k}  \right) = 
\mathcal{G}_{b,k} =\int_0^{\infty}\tfrac{1}{\mu +\alpha_{b,k}}\mathcal{F}_{1/\rho_{b,k}}(\mu) d \mu,
\end{equation} 
where $\mathcal{F}_{\!1/\rho_{b,k}}(\mu)\!\stackrel{\Delta}{=}\!(1\!-\!1/\rho_{b,k})^+\delta(\mu) \!+\!\frac{\sqrt{(\mu-(1-\sqrt{1/\rho_{b,k}})^2)^+((1+\sqrt{1/\rho_{b,k}})^2-\mu)^+}}{2\pi \mu}$, $\delta(\mu)$ is the Dirac impulse function, and $(x)^+ \!\triangleq\! \max \{0,x\}$. Following a similar approach as for $X_{b,k}$, one can show that
\vspace{-0.5em}
\begin{equation}\label{eqp5.8}
\vspace{-0.5em}
{\hat{\bf{g}}}_{b,k,b}^{\rm H}{\bf{\Phi}}_{b,k}{\bf{\Phi}}_{b,k}{\hat{\bf{g}}}_{b,k,b}
\rightarrow \int_0^{\infty}\tfrac{1}{(\mu +\alpha_{b,k})^2}\mathcal{F}_{1/\rho_{b,k}}(\mu)d\mu
\stackrel{(a)}{=} \overline{\mathcal{G}}_{b,k},
\end{equation}
where $(a)$ is due to $\frac{1}{(\mu +\alpha_{b,k})^2}\!=\!-\frac{d}{d\alpha_{b,k}}\frac{1}{(\mu +\alpha_{b,k})}$.
Based on \eqref{eqp5.2}-\eqref{eqp5.8}, we obtain
\vspace{-0.5em}
\begin{equation}\label{eqp5.9}
\vspace{-0.5em}
\|({\bf{F}}_{b,k}^{\rm H}{\bf{F}}_{b,k}+\alpha_{b,k}{\bf{I}}_M)^{-1}{\hat{\bf{g}}}_{b,k,b}\|^2
\rightarrow \frac{ \overline{\mathcal{G}}_{b,k}}{ \left(1+\mathcal{G}_{b,k}\right)^2}.
\end{equation}
Finally, substituting \eqref{eqp5.9} into \eqref{eqp5.1}, Lemma~\ref{lemma2} is proved.

\vspace{-1em}
\subsection{Proof of Proposition~\ref{prop3}} \label{proof3}
When $M, K \rightarrow \infty$ but $\rho_0$ is finite and fixed, we have $\rho_{b}^{-1}= N_{b}^{\mathrm{n}} / M \le \rho_0^{-1}$. Hence, based on \eqref{eq8.7},  the signal power and the intra-cell interference power of user $(b,k)$ are given as
\vspace{-0.5em}
\begin{align} 
\frac{ P_{b,k}^{\mathrm{s,RZF}} } {M E_{b,k} } &\stackrel{(a)}{=} \frac{| X_{b,k}|^2}
{| 1+ X_{b,k}|^2} \lambda_{b,k}^{\mathrm{RZF}} \hat{\beta}_{b,k,b}  
\stackrel{(b)}{\rightarrow} \frac{ \mathcal{G}^2_{b,k} \hat{\beta}_{b,k,b}} { \overline{\mathcal{G}}_{b,k}}, \label{eqp6.1}\\
\frac{P_{b,k,b,l}^{\mathrm{i,RZF}} }{E_{b,l} } 
&\stackrel{(c)}{=} \frac{ M | {\hat{\bf{g}}}_{b,k,b}^{\mathrm H} {\bf{\Phi}}_{b,l(b,k)} {\hat{\bf{g}}}_{b,l,b}|^2}
{\left| 1+ X_{b,l}|^2| 1+ A_{b,l(b,k)}\right|^2} \cdot \frac{ \left(1+\mathcal{G}_{b,l}\right)^2}{ \overline{\mathcal{G}}_{b,l}}
\hat{\beta}_{b,k,b}, \label{eqp6.2}
\end{align}
respectively, where $\!X_{b,l}\!\!=\!{\hat{\bf{g}}}_{b,l,b}^{\rm H} ({\bf{F}}_{b,l(b,l)}^{\rm H}{\bf{F}}_{b,l(b,l)}+\alpha_{b,l}{\bf{I}}_M\!)^{\!-\!1}{\hat{\bf{g}}}_{b,l,b}$, ${\bf{\Phi}}_{b,l(b,k)}=({\bf{F}}_{b,l(b,l,b,k)}^{\rm H}{\bf{F}}_{b,l(b,l,b,k)} +\alpha_{b,l} {\bf{I}}_M)^{-1}$,  and $A_{b,l(b,k)}\!\!=\!{\hat{\bf{g}}}_{b,k,b}^{\rm H} {\bf{\Phi}}_{b,l(b,k)} {\hat{\bf{g}}}_{b,k,b}$. In \eqref{eqp6.1},
$(a)$ is due to the matrix inversion lemma \cite{Nguyen2008}, and $(b)$ follows from \eqref{eqp5.5} and Lemma~\ref{lemma2}. In \eqref{eqp6.2},
$(c)$ follows from the repeated application of the matrix inversion lemma \cite{Nguyen2008}, whereby ${\bf{F}}_{b,l(b,l)}$ is obtained by deleting vector ${\hat{\bf{g}}}_{b,l,b}$ from ${\bf{F}}_{b,l}$ and in ${\bf{F}}_{b,l(b,l,b,k)}$,  vector ${\hat{\bf{g}}}_{b,k,b}$ is further deleted from ${\bf{F}}_{b,l(b,l)}$.
Using \eqref{eqp5.5}, we obtain $X_{b,l}\rightarrow \mathcal{G}_{b,l}$ and $A_{b,l(b,k)}\rightarrow \mathcal{G}_{b,l}$, { {as $M, K\rightarrow \infty$ but $\rho_{b,l}^{-1} = M / N_{b,l}^{\mathrm{n}} \le \rho_0^{-1}$ is finite.}} Additionally, we have
\vspace{-0.5em}
\begin{equation} \label{eqp6.3}
 M \left| {\hat{\bf{g}}}_{b,k,b}^{\mathrm H}{\bf{\Phi}}_{b,l(b,k)}{\hat{\bf{g}}}_{b,l,b}\right|^2
\!=\!\! \lim_{M\to\infty} \tfrac{M}{M} \mathrm{tr}\left({\bf{\Phi}}_{b,l}{\hat{\bf{g}}}_{b,l,b} {\hat{\bf{g}}}_{b,l,b}^{\mathrm H} {\bf{\Phi}}_{b,l}^{\mathrm H}\right) 
\!=\!\! \lim_{M, K\to\infty} {\hat{\bf{g}}}_{b,l,b}^{\mathrm H}{\bf{\Phi}}_{b,l}^{\mathrm H} {\bf{\Phi}}_{b,l}{\hat{\bf{g}}}_{b,l,b}
\stackrel{(d)}{=} \overline{\mathcal{G}}_{b,l},
\end{equation}
where $(d)$ is obtained using a similar approach as in \eqref{eqp5.8} as $M, K \rightarrow \infty$. 

Moreover, the inter-cell interference power, $P_{b,k,j,l}^{\mathrm{i,RZF}},j\neq b$, has to be evaluated for three cases, similar to $P_{b,k,j,l}^{\mathrm{i,ZF}}$ in \eqref{eq7.6}: (i) If $l=k$, as ${\hat{\bf{g}}}_{b,k,j} = {\hat{\bf{g}}}_{j,k,j}$, we have 
\vspace{-0.5em}
\begin{equation} \label{eqp6.4}
\frac{ P_{b,k,j,k}^{\mathrm{i,RZF}} }{M E_{j,k}} 
{=}\frac{| {\hat{\bf{g}}}_{j,k,j}^{\mathrm H}({\bf{F}}_{j,k(j,k)}^{\rm H}{\bf{F}}_{j,k(j,k)}+\alpha_{j,k}{\bf{I}}_M)^{-1}{\hat{\bf{g}}}_{j,k,j} |^2 }{|1+ {\hat{\bf{g}}}_{j,k,j}^{\mathrm H}({\bf{F}}_{i,k(j,k)}^{\rm H}{\bf{F}}_{j,k(j,k)}+\alpha_{j,k}{\bf{I}}_M)^{-1}{\hat{\bf{g}}}_{j,k,j}|^2 } \hat{\beta}_{b,k,j} \lambda_{b,l}^{\mathrm{RZF}} 
{\rightarrow} \frac{\mathcal{G}^2_{j,k} \hat{\beta}_{b,k,j}}{ \overline{\mathcal{G}}_{j,k}},
\end{equation}
similar to \eqref{eqp5.2} and \eqref{eqp5.5}. 
(ii) If $l\neq k$ and ${\hat{\bf{g}}}_{j,k,j}\in {\bf{F}}_{j,l}$, we have ${\hat{\bf{g}}}_{b,k,j}\in {\bf{F}}_{j,l}$ as ${\hat{\bf{g}}}_{j,k,j}={\hat{\bf{g}}}_{b,k,j}$, and 
\vspace{-0.5em}
\begin{equation} \label{eqp6.7b}
\vspace{-0.5em}
\frac{ P_{b,k,j,l}^{\mathrm{i,RZF}} }{E_{j,l}} 
\!\!=\!\! M | {\hat{\bf{g}}}_{j,k,j}^{\mathrm H}({\bf{F}}_{j,l}^{\rm H}{\bf{F}}_{j,l}\!+\!\alpha_{j,l}{\bf{I}}_M)^{\!-\!1}{\hat{\bf{g}}}_{j,l,j} |^2 \hat{\beta}_{b,k,j} \lambda_{b,l}^{\mathrm{RZF}} 
\rightarrow  \frac{ \hat{\beta}_{b,k,j} } {(1\!+\!\mathcal{G}_{j,l})^2}\!,\!\!
\end{equation}
when $M, K\rightarrow \infty$.
(iii) If $l\neq k$ and ${\hat{\bf{g}}}_{j,k,j}\notin {\bf{F}}_{j,l}$, i.e., ${\hat{\bf{g}}}_{b,k,j}$ and ${\bf{F}}_{j,l}$ are independent, we have
\vspace{-0.5em}
\begin{align} \label{eqp6.5}
\frac{ P_{b,k,j,l}^{\mathrm{i,RZF}} }{E_{j,l}} 
{=}\frac{ M | {\hat{\bf{g}}}_{b,k,j}^{\mathrm H}({\bf{F}}_{j,l(j,l)}^{\rm H}{\bf{F}}_{j,l(j,l)}+\alpha_{j,l}{\bf{I}}_M)^{-1}{\hat{\bf{g}}}_{j,l,j} |^2 }{|1+ {\hat{\bf{g}}}_{j,l,j}^{\mathrm H}({\bf{F}}_{j,l(j,l)}^{\rm H}{\bf{F}}_{j,l(j,l)}+\alpha_{j,l}{\bf{I}}_M)^{-1}{\hat{\bf{g}}}_{j,l,j} |^2 } \hat{\beta}_{b,k,j} \lambda_{b,l}^{\mathrm{RZF}}
\stackrel{(h)}{\rightarrow} \hat{\beta}_{b,k,j}.
\end{align}

Finally, as ${\tilde{\bf{h}}}_{b,k,j}$ and ${\bf{F}}_{j,l}$ are independent, a similar approach as in \eqref{eqp6.5} can be employed to show that the interference power caused by the estimation CSI error, $P_{b,k,j,l}^{\mathrm{e,RZF}}$, satisfies
\vspace{-0.5em}
\begin{equation} \label{eqp6.6}
\vspace{-0.5em}
P_{b,k,j,l}^{\mathrm{e,RZF}}={\tilde{\beta}_{b,k,j}}E_{j,l},
\end{equation}
when $M, K \rightarrow \infty$. Substituting \eqref{eqp6.1}--\eqref{eqp6.6} into \eqref{eq8.6}, Proposition~\ref{prop3} is proved.

\vspace{-1em}
\subsection{Proof of Lemma~\ref{lem4}} \label{lemma4}
Since, with coded caching, users with the same index in different cells cache the same set of subfiles for each file, in the following, we only need to analyze the placement of an arbitrary file. In particular, if $l=k$, user $(b,k)$ and user $(j,l)$ cache exactly the same set of subfiles for this file. Hence, the subfiles to be delivered to user $(j,l)$ are not cached at user $(b,k)$, and $p_{k,l}^{\mathrm{i}}=1$. Otherwise, i.e., if $l\neq k$, user $(b,k)$ and user $(j,l)$ both cache $C_{K-1}^{t-1}$ subfiles of this file, where $C_{K-2}^{t-2}$ subfiles are identical. Since each user needs to receive $C_{K}^{t}-C_{K-1}^{t-1}$ subfiles, the probability that user $(b,k)$ is interfered by user $(j,l)$ is $p_{k,l}^{\mathrm{i}}=\frac{(C_{K}^{t}-C_{K-1}^{t-1})-(C_{K-1}^{t-1}-C_{K-2}^{t-2})}{C_{K}^{t}-C_{K-1}^{t-1}}=\frac{K-t-1}{K-1}$, which is independent of the cell index. Similarly, the probability that user $(j,l)$ is interfered by user $(b,k)$ is $p_{k,l}^{\mathrm{n}}=1,l=k$, and $p_{k,l}^{\mathrm{n}}=\frac{(C_{K}^{t}-C_{K-1}^{t-1})-(C_{K-1}^{t-1}-C_{K-2}^{t-2})}{C_{K}^{t}-C_{K-1}^{t-1}}=\frac{K-t-1}{K-1},l\neq k$, which is independent of the user and subfile indices. This completes the proof of Lemma~\ref{lem4}.
\end{appendix}


\vspace{-0.9em}
\begin{thebibliography}{99}
\bibitem{Wei2019}
X. Wei, L. Xiang, L. Cottatellucci, T. Jiang, and R. Schober, ``Cache-aided massive MIMO: Linear precoding design and performance analysis,'' in {\em \it Proc. IEEE ICC\/}, Shanghai, China, May 2019.

\bibitem{Wong2017}
V. W. S. Wong, R. Schober, D. W. K. Ng, and L.-C. Wang, {\em \it Key Technologies for 5G Wireless Systems\/}, Cambridge University Press, 2017.

\bibitem{N2013}
H. Q. Ngo, E. G. Larsson, and T. L. Marzetta, ``Energy and spectral efficiency of very large multiuser MIMO systems,'' {\em \it IEEE Trans. Commun.\/}, vol. 61, no. 4, pp. 1436-1449, Apr. 2013.

\bibitem{Lu2014}
{ {J. Zhang, E. Björnson, M. Matthaiou, D. W. K. Ng, H. Yang, and D. J. Love, ``Prospective multiple antenna technologies for beyond 5G," \emph{IEEE J. Sel. Areas Commun.}, vol. 38, no. 8, pp. 1637-1660, Aug. 2020.}}

\bibitem{Marzetta2010}
T. L. Marzetta, ``Noncooperative cellular wireless with unlimited numbers of base station antennas,'' {\em \it IEEE Trans. Wireless Commun.\/}, vol. 9, no. 11, pp. 3590-3600, Nov. 2010.

\bibitem{Huh2011}
H. Huh, S. Moon, Y. Kim, I. Lee, and G. Caire, ``Multi-cell MIMO downlink with cell cooperation and fair scheduling: A largesystem limit analysis,'' {\em \it IEEE Trans. Inf. Theory\/}, vol. 57, no. 12, pp. 7771-7786, Dec. 2011.


\bibitem{Yang2013}
H. Yang and T. L. Marzetta, ``Performance of conjugate and zeroforcing beamforming in large-scale antenna systems,'' {\em \it IEEE J. Sel. Areas Commun.\/}, vol. 31, no. 2, pp. 172-179, Feb. 2013.


\bibitem{Ngo2013}
H. Q. Ngo, E. G. Larsson, and T. L. Marzetta, ``Massive MU-MIMO downlink TDD systems with linear precoding and downlink pilots,'' {\em \it in Proc. Allerton Conf. Commun. Control Comput.\/}, Monticello, IL, USA, Oct. 2013.

\bibitem{Raeesi2018}
O. Raeesi, A. Gokceoglu, Y. Zou, E. Bj\"ornson, and M. Valkama, ``Performance analysis of multi-user massive MIMO downlink under channel non-reciprocity and imperfect CSI,'' {\em \it IEEE Trans. Commun.\/}, vol. 66, no. 6, pp. 2456-2471, June 2018.


\bibitem{Hoydis2013}
J. Hoydis, S. ten Brink, and M. Debbah, ``Massive MIMO in the UL/DL of cellular networks: How many antennas do we need?'' {\em \it IEEE J. Sel. Areas Commun.\/}, vol. 31, no. 2, pp. 160-171, Feb. 2013.


\bibitem{Muller2014}
R. M\"uller,  L. Cottatellucci, and  M. Vehkaper\"a, ``Blind pilot decontamination,'' {\em \it IEEE J. Sel. Areas Commun.\/}, vol. 8, no. 5, pp. 773-786, Oct. 2014.

\bibitem{Yu2017}
W. Yu, ``On the fundamental limits of massive connectivity,'' {\em \it in Proc. Inf. Theory App. (ITA) Workshop\/}, San Diego, CA, Feb. 2017.

\bibitem{Emil2019}
E. Bj\"ornson, L. Sanguinetti, H. Wymeersch, et al., ``Massive MIMO is a reality --- What is next? Five promising research directions for antenna arrays", \emph{Digital Signal Process.},   vol.~94, pp. 3-20, 2019.



\bibitem{Semiari2018}
O. Semiari, W. Saad, and M. Bennis, ``Caching meets millimeter wave communications for enhanced mobility management in 5G networks,'' {\em \it IEEE Trans. Wireless Commun.\/}, vol. 17, no. 2, pp. 779-793, Feb. 2018.

\bibitem{Ng2018}
L. Xiang, D. W. K. Ng, R. Schober, and V. W. S. Wong, ``Cache-enabled physical layer security for video streaming in backhaul-limited cellular networks,'' {\em \it IEEE Trans. Wireless Commun.\/}, vol. 17, no. 2, pp. 736-751, Feb. 2018.

\bibitem{Maddah-Ali2014}
M. A. Maddah-Ali and U. Niesen, ``Fundamental limits of caching'' {\em \it IEEE Trans. Inf. Theory\/}, vol. 60, no. 5, pp. 2856-2867,
May 2014.

\bibitem{TWC22a}
{ {H. Zhao, A. Bazco-Nogueras, and P. Elia, ``Wireless coded caching can overcome the worst-user bottleneck by exploiting finite file sizes," \emph{IEEE Trans. Wireless Commun.}, to appear, 2022.}}

\bibitem{TWC22b}
{ {Y. Liu, A. Tang, and X. Wang, ``Joint scheduling and power optimization for delay constrained transmissions in coded caching over wireless fading channels," \emph{IEEE Trans. Wireless Commun.}, vol. 21, no. 3, pp. 2093-2106, Mar. 2022.}}

\bibitem{Xiang2018}
%
L. Xiang, D. W. K. Ng, X. Ge, Z. Ding, V. W. S. Wong, and R. Schober, ``Cache-aided non-orthogonal multiple access: The two-user case,'' {\em \it IEEE J. Sel. Topics Signal Process.\/}, vol. 13, no. 3, pp. 436-451, Jun. 2019.

\bibitem{Dani2021}
M. N. Dani, D. K. C. So, J. Tang, and Z. Ding, ``NOMA and coded multicasting in cache-aided wireless networks," \emph{IEEE Trans. Wireless Commun.}, vol. 21, no. 4, to appear, Apr. 2022. 

\bibitem{Jaafar2020}
W. Jaafar, S. Naser, S. Muhaidat, P. C. Sofotasios, and H. Yanikomeroglu, ``Multiple access in aerial networks: From orthogonal and non-orthogonal to rate-splitting," \emph{IEEE Open J. Veh. Techno.}, vol. 1, pp. 372-392, 2020

\bibitem{Ngo2018}
K. H. Ngo, S. Yang, and M. Kobayashi, ``Scalable content delivery with coded caching in multi-antenna fading channels,'' {\em \it IEEE Trans. Wireless Commun.\/}, vol. 17, no. 1, pp. 548-562, Jan. 2018.

\bibitem{Papazafeiropoulos2018}
A. Papazafeiropoulos and T. Ratnarajah, ``Modeling and performance of uplink cache-enabled massive MIMO heterogeneous networks,'' {\em \it IEEE Trans. Wireless Commun.\/}, vol. 17, no. 12, pp. 8136-8149, Dec. 2018.

\bibitem{Zhu2016}
J. Zhu, R. Schober, and V. K. Bhargava, ``Linear precoding of data and artificial noise in secure massive MIMO systems,'' {\em \it
IEEE Trans. Wireless Commun.\/}, vol. 15, no. 3, pp. 2245-2261, Mar. 2016.


\bibitem{Kay1993}
S. Kay, {\em \it Fundamentals of Statistical Signal Processing: Estimation
Theory\/}, Prentice Hall, 1993.

\bibitem{TM2016}
T. L. Marzetta, E. G. Larsson, H. Yang, and H. Q. Ngo, {\em \it Fundamentals of Massive MIMO}, Cambridge University Press, 2016.

\bibitem{Access21}
{ {M. A. Albreem, A. H. A. Habbash, A. M. Abu-Hudrouss, and S. S. Ikki, ``Overview of precoding techniques for massive MIMO," \emph{IEEE Access}, vol. 9, pp. 60764-60801, 2021.}}


\bibitem{Chien2016}
T. V. Chien, E. Bj\"ornson, and E. G. Larsson, ``Joint power allocation and user association optimization for massive MIMO systems,'' {\em \it IEEE Trans. Wireless Commun.\/}, vol. 15, no. 9, pp. 6384-6399, Sep. 2016.


\bibitem{Sifaou2014}
H. Sifaou, A. Kammoun, L. Sanguinetti, M. Debbah, and M-S. Alouini, ``Power efficient low complexity precoding for massive MIMO systems,'' {\em \it in Proc. IEEE GlobalSIP\/}, Atlanta, GA, USA, Dec. 2014.

\bibitem{Nguyen2008}
V. K. Nguyen and J. S. Evans, ``Multiuser transmit beamforming via regularized channel inversion: A large system analysis,'' {\em \it in Proc. IEEE Global Commun. Conf.\/}, New Orleans, LA, USA, Dec. 2008.

\bibitem{Malak2014}
{ {N. Garg, M. Sellathurai, V. Bhatia, and T. Ratnarajah, ``Function approximation based reinforcement learning for edge caching in massive MIMO networks," \emph{IEEE Trans. Commun.}, vol. 69, no. 4, pp. 2304-2316, Apr. 2021.}}

\bibitem{Blaszczyszyn2015}
{ {K. Wang, W. Chen, J. Li, Y. Yang, and L. Hanzo, ``Joint task offloading and caching for massive MIMO-aided multi-tier computing networks," \emph{IEEE Trans. Commun.}, vol. 70, no. 3, pp. 1820-1833, Mar. 2022.}}


\bibitem{Kokoska2000}
S. Kokoska and D. Zwillinger, {\em \it Standard Probability and Statistics Tables and Formulae\/}, FL, Boca Raton: CRC Press, 2000.


\bibitem{Smythe1973}
R. T. Smythe, ``Strong laws of large numbers for $\gamma$-dimensional arrays of random variables,''
{\em \it The Annals of Probability\/}, vol. 1, no. 1, pp. 164-170, Feb. 1973.


\bibitem{Adhikary2013}
A. Adhikary, J. Nam, J.-Y. Ahn, and G. Caire, ``Joint spatial division and multiplexing: The large-scale array regime,'' {\em \it IEEE Trans. Inf. Theory\/}, vol. 59, no. 10, pp. 6441-6463, Oct. 2013.

\bibitem{Wen2015}
C. Wen, S. Jin, K. Wong, J. Chen, and P. Ting, ``Channel estimation for massive MIMO using Gaussian-mixture Bayesian learning,'' {\em \it IEEE Trans. Wireless Commun.\/}, vol. 14, no. 3, pp. 1356-1368, Mar. 2015.

%
%

\bibitem{Evans2000}
J. S. Evans and D. N. C. Tse, ``Large system performance of linear multiuser receivers in multipath fading channels,'' {\em \it IEEE Trans. Inf. Theory\/}, vol. 46, pp. 2059-2078, Sept. 2000.



\end{thebibliography}
\end{document}